\documentclass[manuscript,11pt]{aastex}
\usepackage{epsfig}

\def\fei{Fe\,{\sc i}}

\def\fexvii{Fe\,{\sc xvii}}

\def\fexxv{Fe\,{\sc xxv}}
\def\fexxvi{Fe\,{\sc xxvi}}

\def\mathv{\textbf{\em v}}
\def\mathvphi{\textbf{\em v}_{\phi}}
\def\mathvp{\textbf{\em v}_{p}}
\def\mathB{\textbf{\em B}}
\def\mathBp{\textbf{\em B}_p}
\def\mathBphi{\textbf{\em B}_{\phi}}

\def\mathJ{\textbf{\em J}}

\def\mathphihat{\mbox{\boldmath$\hat{\phi}$}}

\def\cm{\ifmmode {\rm cm}^{-1} \else cm$^{-1}$ \fi}
\def\s{\ifmmode {\rm s}^{-1} \else s$^{-1}$ \fi}
\def\cc{\ifmmode {\rm cm}^{-3} \else cm$^{-3}$ \fi}
\def\cs{\ifmmode {\rm cm}^{-2} \else cm$^{-2}$ \fi}
\def\g{\ifmmode \gamma \else $\gamma$\fi}
\def\G{\ifmmode \Gamma \else $\Gamma$\fi}
\def\Gs{\ifmmode \Gamma~ \else $\Gamma~$\fi}

\def\gc{\ifmmode \gamma_{\rm c} \else $\gamma_{\rm c}$ \fi}
\def\sw{Schwarzschild~}
\def\gsim{\mathrel{\raise.5ex\hbox{$>$}\mkern-14mu
             \lower0.6ex\hbox{$\sim$}}}
\def\lsim{\mathrel{\raise.3ex\hbox{$<$}\mkern-14mu
             \lower0.6ex\hbox{$\sim$}}}
\def\simless{\mathbin{\lower 3pt\hbox
     {$\rlap{\raise 5pt\hbox{$\char'074$}}\mathchar"7218$}}}   
\def\simmore{\mathbin{\lower 3pt\hbox
     {$\rlap{\raise 5pt\hbox{$\char'076$}}\mathchar"7218$}}}   
\def\Msun{M_\odot}                                
\def\4u{4U 1728--34}
\def\deg{^\circ}

\newcommand{\Alfven}{Alfv$\acute{\rm e}$n}
\newcommand{\Alfvenic}{Alfv$\acute{\rm e}$nic}

\lefthead{et al.} \righthead{\4u}

\shorttitle{Disk-Winds and Jets}

\shortauthors{Fukumura et al.}

\begin{document}

\title{Stratified Magnetically-Driven Accretion-Disk Winds and Their Relations to Jets }

\date{\today}

\author{\textsc{Keigo Fukumura}\altaffilmark{1,2,3},
\textsc{Francesco Tombesi}\altaffilmark{3,4}
\textsc{Demosthenes Kazanas}\altaffilmark{3},
\textsc{Chris Shrader}\altaffilmark{3,5},
\textsc{Ehud Behar}\altaffilmark{6},
\textsc{and}
\textsc{Ioannis Contopoulos}\altaffilmark{7} }

\altaffiltext{1}{Email: fukumukx@jmu.edu}
\altaffiltext{2}{University of Maryland, Baltimore County
(UMBC/CRESST), Baltimore, MD 21250} \altaffiltext{3}{Astrophysics
Science Division, NASA/Goddard Space Flight Center, Greenbelt, MD
20771} \altaffiltext{4}{Department of Astronomy, University of Maryland, College Park, MD 20742} \altaffiltext{5}{University Space Research Association, 10211 Wincopin Circle, Suite 620, Columbia, MD 21044} \altaffiltext{6}{Department of
Physics, Technion, Haifa 32000, Israel} \altaffiltext{7}{Research Center for Astronomy, Academy of
Athens, Athens 11527, Greece}

\begin{abstract}

\baselineskip=15pt

%

We explore the poloidal structure of two-dimensional (2D) MHD winds in
relation to their potential association with the X-ray warm
absorbers (WAs) and the highly-ionized ultra-fast outflows (UFOs) in
AGN, in a single unifying approach. We present the density
$n(r,\theta)$, ionization parameter $\xi(r,\theta)$, and velocity
structure $v(r,\theta)$ of such ionized winds for typical values of
their fluid-to-magnetic flux ratio, $F$, and specific angular
momentum, $H$, for which wind solutions become super-\Alfvenic. We
explore the geometrical shape of winds for different values of these
parameters and delineate the values that produce the widest and
narrowest opening angles of these winds, quantities necessary in the
determination of the statistics of AGN obscuration. We find that
winds with smaller $H$ show a poloidal geometry of narrower opening
angles with their \Alfven\ surface at lower inclination angles and
therefore they produce the highest line of sight (LoS) velocities for
observers at higher latitudes with the respect to the disk plane.
We further note a physical and spatial correlation between the X-ray
WAs and UFOs that form along the same LoS to the observer but at
different radii, $r$, and distinct values of $n$, $\xi$ and $v$
consistent with the latest spectroscopic data of radio-quiet Seyfert galaxies.
We also show that, at least in the case of 3C 111, the winds'
pressure is sufficient to contain the
relativistic plasma responsible for its radio emission. Stratified
MHD disk-winds could therefore serve as a unique means to
understand and unify the diverse AGN outflows.


\end{abstract}


\baselineskip=15pt

\section{Introduction}

Outflows are a common occurrence in accretion-powered objects. Their
presence is affirmed by blue-shifted absorption features in their
optical, UV and X-ray spectra at frequencies of well defined
transitions. Their velocities span a range of $10^3$, roughly
between 100 and 100,000 km/s, while the broad ($\sim 10,000$ km/s)
absorption troughs of UV transitions are the defining characteristic
of the class of broad absorption line quasars (BAL QSOs). The launch of {\it Hubble, ASCA, Chandra},
{\it XMM-Newton} and {\it Suzaku} and their superior sensitivity,
resolution and energy bandwidth showed that $\simeq 50\%$ of AGN
exhibit signatures of such outflows in their optical, UV and/or
X-ray spectra. Of these the high resolution long X-ray observations
of {\em Chandra} and {\em XMM-Newton} are of particular interest
because they discovered transitions that span a range of $\sim 10^5$
in ionization parameter\footnote[1]{$\xi \equiv L/(n r^2)$ where $L$
is an ionizing luminosity (usually defined between 1 and 1000 Ryd),
$n$ is the plasma number density and $r$ is distance from the
ionizing source.}, $\xi$, (such as \fei\ through \fexxvi\, among
others) and therefore sample a very broad range of conditions
for the column density and velocity of plasma along the
observers' line of sight (LoS). The fact that the plethora of these
transitions spanning 5 decades in $\xi$ is squeezed within
roughly 1.5 decades of frequency, underscores the utility of X-ray
spectroscopy.

The first signature of X-ray absorbing plasma in AGN was that in the
{\em Einstein} spectrum of the QSO MR 2251-178 \citep{Halpern84},
attributed to ``warm" plasma of temperature $T \sim 10^6$ K (rather
than ``cool" $T \simeq 10^4$ K clouds), thus coining the term warm
absorber (WA) for absorption features in the $\lsim 1$ keV band. The
ubiquity of WAs was eventually established by {\em ASCA} which
discovered absorption features of typical column density of $N_H
\sim 10^{20}-10^{22}$ cm$^{-2}$ and $\xi \sim 10^{-1}-10^4$
erg~cm~s$^{-1}$ at moderate outflow velocities ($v \lesssim 3,000$
km~s$^{-1}$) in the spectra of $\sim 50\%$ of Seyfert 1's
\citep{Reynolds97,George98}. Following these discoveries, extensive
spectroscopic observations with {\em Chandra} and {\em XMM-Newton} have
been made to study in detail the physical conditions and spatial
origin of WAs
\citep[e.g.][]{Blustin05,McKernan07,Reeves09a,Turner09,Torresi10,Torresi12}.
These are of interest because the presence of ionized absorbing gas
of a very wide range of $\xi$ along the observers' LoS, offers the
opportunity of mapping the spatial distribution of this gas: one can
easily see that the knowledge of $\xi$ (from the presence of certain
ionic species) and the measurement of their hydrogen equivalent
column $N_H \;[\lesssim n(r) r$] in the observed spectrum can provide
a measure of the absorbing gas' density dependence on the distance
from the continuum source along the observer's LoS. Such information
is extremely valuable in assessing the global properties of the
winds which presumably give rise to the observed absorbers.

A quantitative, more systematic formal way of implementing the above
procedure is to construct the absorption measure distribution (AMD),
namely the hydrogen-equivalent column of specific ions per decade of
$\xi$; i.e. ${\rm AMD} \equiv d N_{\rm H}/d \log \xi$
\citep[e.g.][among others]{HBK07,B09,Detmers11,Holczer12}. These
authors, instead of adding gas components at different values of
$\xi$ and $N_H$ until a satisfactory statistical significance $\chi^2$ is obtained, they
assumed a continuous dependence of binned $N_H$ on $\xi$ (i.e. $d
N_H/d \log \xi \propto \xi^{s}$) and provided a global fit to the
entire set of ionic transitions with $s$ as the sole parameter. The
continuous dependence of $N_H$ on $\xi$ implies (assuming a smooth
spatial gas distribution) also a continuous dependence of the
ionized medium density with $r$, its distance from the AGNs, of the
form $n(r) \propto r^{-(2s+1)/(s+1)}$. X-ray analysis of the spectra
from a number of Seyfert galaxies showed that $N_H$ has only a weak
dependence on $\xi$ ($s \simeq 0$) \citep{B09} implying a wind
density profile of $n(r) \propto r^{-1}$ (the largest value of $s$
found was $s \simeq 0.3$ implying $n(r) \propto r^{-1.25}$).
Furthermore, the presence of UV absorption features in the spectra
of AGNs that exhibit X-ray WAs \citep[e.g.][]{Crenshaw03} suggests a
physical link between these components, which at present is not
completely understood.

In parallel with the high resolution observations discussed above, a
number of lower resolution CCD X-ray observations of radio-quiet
Seyferts and BAL QSOs have
discovered absorption features identified as highly-ionized iron
(primarily \fexxv /\fexxvi\ at energies $\sim 7-8$ keV in the source
frame) of column densities ($N_H \sim 10^{23}-10^{24}$ cm$^{-2}$),
generally higher than those of their moderately-ionized ions at
higher (blueshifted) velocities ($v/c \sim 0.1-0.7$ where $c$ is the
speed of light), coined ultra-fast outflows (UFOs;
\citealt{Tombesi10a}). These are seen across different AGN
populations such as Seyferts, BAL QSOs and non-BAL QSOs
\citep[][]{Chartas03,Chartas07,Reeves09b,Pounds06,Tombesi10a,Tombesi11a,Tombesi12a}.
It is worth noting that similar features (UFOs) have also been
reported in radio-loud Seyferts such as 3C~111
\citep[e.g.][]{Tombesi10b,Tombesi11b}, indicating that these high
outflow velocity transitions  represent a generic feature of AGN
rather than one associated with a specific AGN subclass.
Furthermore, in the case of a BAL QSO APM~08279+5255,
\citet{Chartas09} have noted a likely correlation between X-ray
photon index $\Gamma$ and the measured outflow velocity of \fexxv,
implying additional underlying physics pertaining to the velocities
of these absorbers.

These observations of high ionization ($\log \xi \gtrsim 4$), high
velocity ($v \gsim 0.1 c$) outflowing gas (UFOs) are of importance
because they challenge the conventional radiative acceleration
scenario of AGN outflows \citep[e.g.][]{PSK00} which demands low
ionization for this gas so that line radiation pressure driving be
efficient; as such, they lend support to magnetic driving for these
flows.
The recent discovery of the UFOs in radio-loud Seyferts exhibiting
jets \citep[][]{Tombesi10b,Tombesi11b} seems also to challenge a
simplistic {\it wind-jet dichotomy} that associates ``winds" with
radio quiet AGN and ``jets" with the radio loud ones, perhaps
implying a multi-component gas in the same outflowing plasma across
these broad AGN classes. With little simultaneous X-ray-radio
studies having been conducted so far in radio-loud AGNs,
\citet[][hereafter T12b]{Tombesi12b} did perform  such a combined
study of the X-ray UFOs and the VLBA radio jet of the radio galaxy
3C~111; their simultaneous presence in this object has indicated
that these components can actually coexist in AGN.  Also, with
respect to the coexistence of WAs and UFOs in radio-quiet Seyferts,
\citep[e.g.][]{Tombesi10a,Tombesi11a,Tombesi12a}, \citet[][hereafter
T12c]{Tombesi12c} have analyzed X-ray data from 35 radio-quiet
Seyferts to obtain a number of correlations between $\xi$, $N_H$ and
$v$  of these components that argue in support of a single physical
entity underlying the nature of WAs and UFOs.

Similar absorption features have been observed and studied
intensively in binary systems including transient black hole (BH)
candidates and neutron star low-mass X-ray binaries (LMXBs). These
exhibit typically highly ionized plasma (e.g. Si, S, Mg, Ar and Fe)
of $\log \xi \sim 3-4$ and $N_H \sim 10^{21}-10^{22}$ cm$^{-2}$ at
typical outflow velocities ranging from $v \sim 300$ km~s$^{-1}$ to
$2,000$ km~s$^{-1}$, i.e. properties similar to those of AGN but
with lower velocities and more highly ionized species. Recently,
however, \citet{King12} found an extremely fast ($v/c \sim
0.03-0.05$) ionized outflow based on the \fexxv\ K-shell signature
from {\it Suzaku}/XIS observations of a transient X-ray source,
IGR~J17091-3624. This is currently the fastest X-ray wind known in
galactic binaries. This clearly illustrates a broad range of X-ray
outflow kinematics possible around stellar-size compact objects.
Some of the BH binaries, such as GRS~1915+105
\citep[e.g.][]{Lee09,Neilsen12} and H~1743-322
\citep[e.g.][]{Miller06,Blum10}, exhibit relativistic radio jets
akin to those of radio-loud Seyferts. In particular these two
components of outflows, X-ray absorbers and radio jets, appear to be
anti-correlated: X-ray winds seem to be suppressed in the low/hard
state of these objects  which favor the presence of radio jets
\citep[e.g.][]{Fender04}. Also, VLBA monitoring of jet emission
combined with simultaneous {\it RXTE} observations of low-frequency
quasi-periodic oscillations (LFQPOs) seems to imply that the jet
ejection is anti-correlated with the disk activity \citep[see,
e.g.,][for utilizing QPOs as a proxy of disk activity in
H~1743--322]{Miller-Jones12}. These observations are suggestive of a
mutual interplay between the disk (where X-ray outflows are also
launched) and jet production.

In an effort to place the ensemble of these facts into the broader
unifying context of accretion flows onto compact objects and their
associated winds, \citet[][hereafter FKCB10a]{FKCB10a} have
considered the photoionization of magnetohydrodynamic (MHD)
accretion-disk winds. Motivated by the observed broad range of the
WA and  UFO ionization parameter and velocity range, they opted for
self-similar models of MHD winds, as they naturally cover many
decades in radius. Since their enunciation by \citet[][hereafter, BP82]{BP82}, these
winds have been the subject of many studies both semi-analytical
(BP82; \citealt{CL94}, hereafter CL94; \citealt{KK94}, hereafter KK94;
\citealt{Everett05}; \citealt{Ferreira97}) and fully numerical
\citep[][]{Proga03,Pudritz06,Fendt06,Ohsuga09,PorthFendt10,Murphy10}.
%
%
A crucial hint for the choice of a specific model has been the value
of the parameter $s(\simeq 0)$ in the AMD relation (the dependence
of $N_H$ on $\xi$), since this determines the run of the wind
density with distance. Clearly, observations favor density profiles
close to $n(r) \propto 1/r$, a value significantly different from
that of the winds of BP82 (which imply $s \simeq 3/2$) and at
significant odds with radiation driven outlflows (which, for $\xi$
approaching its asymptotic value $\xi_{\infty}$, imply $s \gg 1$ and
hence $n(r) \propto r^{-2}$). The important features of these
density profiles is their almost constant (or slowly-decreasing) column density per decade of radius and their linear
decrease of $\xi$ with distance $r$, a fact that determines that
high ionization ions have higher velocities than the lower
ionization ones, in agreement with observation. The models of
FKCB10a reproduce these features yielding velocities $v \sim 300$
km~s$^{-1}$ for the moderately ionized species of \fexvii\ and $v
\sim 3,000$ km~s$^{-1}$ for the highly ionized ions of \fexxvi. This
work was extended by \citet[][hereafter FKCB10b]{FKCB10b}, which
demonstrated the model's broad applicability by successfully
modeling  the combined X-ray UFOs and UV broad absorption features
of the BAL QSO APM~08279+5255 \citep{Chartas09}; this was achieved
by adjusting only the relative normalization of the UV and X-ray
fluxes to those appropriate for BAL QSOs.

Our to date publications of the subject (FKCB10a/b),
presenting the first attempt in this specific direction, have
employed a single set of wind parameters 
implying an identical magnetic field geometry in both cases.
The goal of the present work is to remedy this deficiency by an
exploration of the wind geometry and ionization through a parameter
search, having in mind the X-ray and radio AGN observations
(T12b; T12c), with the underlying theoretical framework still based
on the work of CL94. In \S 2 we summarize the basics of MHD
accretion-disk wind models that has been previously applied for
comparison with the observed X-ray WAs (FKCB10a) and UFOs (FKCB10b).
In \S 3 we present our results and demonstrate a relevance of the
outflows in a number of situations where those winds could provide a
substantial impact on the surrounding environment, and we discuss
the implications of our results in \S 4 with a comparison with BP82 winds in the  Appendix.

\section{Characteristics of Magnetized Accretion-Disk Winds}

As well known, the structure of axisymmetric MHD flows is determined
by the magnetic flux function $\Psi(r,\theta)$ (see e.g. CL94,
\citealt{Ferreira97,VK03,Fendt06,PorthFendt10}). As discussed earlier, the
broad range of the observed $\xi$ and $v$ of absorption features
suggests a power law form of all physical quantities with the radial
coordinate $r$; as a result we assume for $\Psi(r,\theta)$ the
following separation of variables $\Psi(r,\theta)$ in the power-law
form of spherical radius $r$ as $\Psi(r,\theta) \equiv (r/r_o)^q
\tilde{\Psi}(\theta) \Psi_o$ where $q$ is a self-similar index and
$\Psi_o$ is the magnetic flux normalization at innermost radius
$r_o$ at which our scaling is assumed to be valid and
$\tilde{\Psi}(\theta)$ the angular dependence of this function that
has to be determined numerically. We provisionally assume $r=r_o
\equiv 3R_S$ ($R_S$ is the Schwarzschild radius), the value of the
innermost stable circular orbit (ISCO) of the \sw geometry. We make
this choice based on the high velocities ($v \simeq 0.5 c$) observed
in certain objects, with the understanding that those of our results
pertaining to these innermost radii may not be accurate. However, we
do not expect this assumption to affect the wind structure at the
much larger scales covered by the observations.

With the above choice for $\Psi$ the magnetic field strength,
velocity, number density and total pressure take the following form
\begin{eqnarray}
|\mathB(r,\theta)| &\equiv& B_o \left(\frac{r}{r_o}\right)^{q-2} \tilde{B}(\theta)  \ ,
\label{eq:B} \\
|\mathv(r,\theta)| &\equiv& v_o \left(\frac{r}{r_o}\right)^{-1/2} \tilde{v}(\theta) \ ,
\label{eq:v} \\
n(r,\theta) &\equiv& \frac{1}{m_p} \left(\frac{B_o}{v_o}\right)^2
\left(\frac{r}{r_o}\right)^{2q-3} \tilde{n}(\theta) \ , \label{eq:n}  \\
p(r,\theta) &\equiv& K B_o^2 \left(\frac{r}{r_o}\right)^{2q-4} \tilde{n}(\theta)^\Gamma  \ ,
\label{eq:p}
\end{eqnarray}
where
\begin{eqnarray}
\tilde{n}(\theta) &\equiv& \frac{F_o}{4\pi}
\frac{\tilde{B}_p(\theta)}{\tilde{v}_p(\theta)} \ ,  \label{eq:n2}
\end{eqnarray}
and $m_p$ is proton mass and $K$ is dimensionless polytropic
constant related to plasma entropy. All the dimensionless
angular-dependent functions denoted by {\it tilde} must be
numerically obtained from the conservation equations as the solution
to the Grad-Shafranov equation with initial values on the disk at
$(r,\theta) = (r_o,90\deg)$ to which all the quantities are
normalized as denoted by the subscript ``o".
%
%
The magnetized wind is also characterized by its plasma $\beta$
defined as the ratio of thermal to magnetic field pressures
\begin{eqnarray}
\beta(r,\theta) &\equiv& \beta(\theta) =
\frac{p(r,\theta)/K}{B(r,\theta)^2} =
\frac{\tilde{n}(\theta)^\Gamma}{\tilde{B}(\theta)^2} \ ,
\label{eq:beta}
\end{eqnarray}
where $\Gamma=5/3$ is the polytropic index of the wind plasma.
We note that its LoS radial-dependence drops out in our
framework since $B(r) \propto r^{q-2}$ and $p(r) \propto r^{2q-4}$.

Under steady-state axisymmetric conditions there are five conserved
quantities along a streamline of given $\Psi(r,\theta)$; i.e. the
particle flux to magnetic flux ratio $F$, the angular velocity of
field lines $\Omega$, the total angular momentum of the plasma $H$,
the total energy (Bernoulli function) $J$ and the entropy $S$ (see
CL94 for details). $F$ and $H$ are defined as
\begin{eqnarray}
F(\Psi) &\equiv& 4 \pi n m_p \frac{|\mathvp|}{|\mathBp|} \ \label{eq:F} , \\
H(\Psi) &\equiv& -F(\Psi) r v_\phi + r B_\phi \ ,
\end{eqnarray}
where $\mathvp$ and $\mathBp$ are respectively poloidal velocity and
magnetic field while $v_\phi$ and $B_\phi$ are respectively
azimuthal velocity component and magnetic field strength.

We assume the disk threaded by the magnetic field to be infinitely
thin, so we define boundary conditions at the point $(r= r_o, \theta
= 90^{\circ})$ where $\mathv \equiv \mathv_o \sim \mathvphi$ with
only a small component $v_{z,o}$ perpendicular to the disk surface
[$v_r(90\degr), v_\theta(90\degr) \ll v_\phi$]. For a given set
of $F_o$ and $H_o$ a trans-\Alfven\ wind solution can be obtained when the
initial magnetic field line orientation $\theta_o$ is met with the
regularity condition\footnote[2]{While physically valid winds must
pass through a slow magnetosonic point, we assume that the wind
quickly becomes super-slow magnetosonic due to its efficient
acceleration (see CL94 and \citealt{FP95} in relation to the wind conditions).}. With
these constraints,
%
$\Omega_o$ and $J_o$ are dependent quantities constrained by $F_o$
and $H_o$ (see CL94 and FKCB10a); i.e.
\begin{eqnarray}
\Omega_o &=& 1 - v_\theta (90\degr) (F_o + H_o)  \ , \label{eq:Omega}  
\\
J_o &=& v^2_\theta(90\degr) [1 + {\psi (90\degr)}2]/2 - 1/2 - \Omega_o \ , \label{eq:J}
%
\end{eqnarray}
where $\psi(90\degr)$ characterizes the angle the poloidal field/flow lines make with the disk at their footpoints (readers can easily verify that this angle is equal to $\tan^{-1} \{q/\psi(90\degr)\}$ with $v\phi(90\degr) = 1$ and  $v_\theta (90\degr) = 0.01$ at the launching radius $r_o$ (see CL94 for details).
We thus find that, for a given $q$, the wind structure is primarily
governed by $F_o$ and $H_o$; the goal of the present study is to
determine the structure and observable properties (i.e. column
density, LoS velocity as a function of angle) of these winds and
their relation to observations. Specifically, the value of $F_o$ is
generally responsible for determining the wind kinematics while
$H_o$ plays an important role in shaping its poloidal structure.
Note that one must adjust either $F_o$ or $H_o$ for a given initial
poloidal angle of the magnetic field in order for winds to become
super-\Alfvenic\ (see the Appendix for discussion on viability for our choice
of parameter sets). Although the other model parameters (see CL94
and FKCB10a) are also coupled to and affect the wind properties, we
find them to be fundamentally less significant compared to $F_o$ and
$H_o$.

The magnetic and velocity fields can be decomposed
explicitly in poloidal and toroidal components as $\mathB(r,\theta)
\equiv \mathBp + \mathBphi$ and $\mathv(r,\theta) \equiv r \sin
\theta \Omega \mathphihat + m_p F/(4 \pi n) \mathB$, while the
density at ($r_o,90\deg$) can be explicitly expressed in terms of
the dimensionless mass-accretion rate $\dot{m}$ (see FKCB10a) as $
n_o \equiv n(r_o,90\deg) \equiv \eta_W \dot m / (2\sigma_T r_s)$
where $\eta_W$ is the ratio of the total mass flux rate launched in
the wind to $\dot{m}$ (while the rest is accreted) and $\sigma_T$ is
the Thomson cross-section. Note also that in our discussion on
density normalization that $\dot m$ always refers to the local mass
flux at the innermost flow radius at $r \simeq r_o$ because
the wind mass flux generally has a radial dependence. One can also
calculate force elements  in the wind (e.g. gas pressure force and
magnetic forces) as $\nabla p_{\rm wind} \equiv \nabla p_{\rm gas} -
(\mathJ \times \mathB)/c$ (where \mathJ\ is the current density of
the wind) that can be a measure of thermal equilibrium with ambient
gas.

At this point we would like to discuss briefly the issue of
critical points in our winds. These are discussed in BP82, CL94 and
also in \citet{VTST00,FC04}. A complete description would require
that these solutions pass the fast and modified fast (i.e. the
critical point in the direction perpendicular to that of the imposed
symmetry) points as described in BP82 (and achieved in
\citep{VTST00,FC04}). However, these points lie at sufficiently high
latitudes where the wind densities are very low and the gas
ionization complete, so that they are effectively irrelevant to the
absorption features observations. As a result, we limit our wind
description to crossing the \Alfven\ and the slow-magnetosonic
points. While crossing all critical points is an issue of principle
with steady-state winds (e.g. BP82 and CL94 {\it self-similar} winds recollimate on the
axis and even turn back), these
points are in fact crossed without much trouble in nature; i.e. in time-dependent
simulations \citep{Pudritz06,Fendt06,PorthFendt10}. Once the fluid has achieved
escape velocity it reaches infinity (albeit after a potential
recollimation along the symmetry axis).

In the following calculations discussed in \S 3, therefore, we chose to stop the wind solutions past the \Alfven\ point arbitrarily to eliminate physically less reliable portion of it. However, since our primary goal has only been to demonstrate the model's global viability with the observed X-ray absorbers and not to construct a comprehensive model of ultra-relativistic winds/jets, this approximation is justified up to near the \Alfven\ point. Most notably, we find that this manipulation makes little impact on our argument of WAs and UFOs in the context of MHD-driven outflows presented in this paper since most absorbers still lie within such an artificial cut-off.




\begin{figure}[ht]
\begin{center}$
\begin{array}{cc}
\includegraphics[trim=0in 0in 0in
0in,keepaspectratio=false,width=3.9in,angle=-0,clip=false]{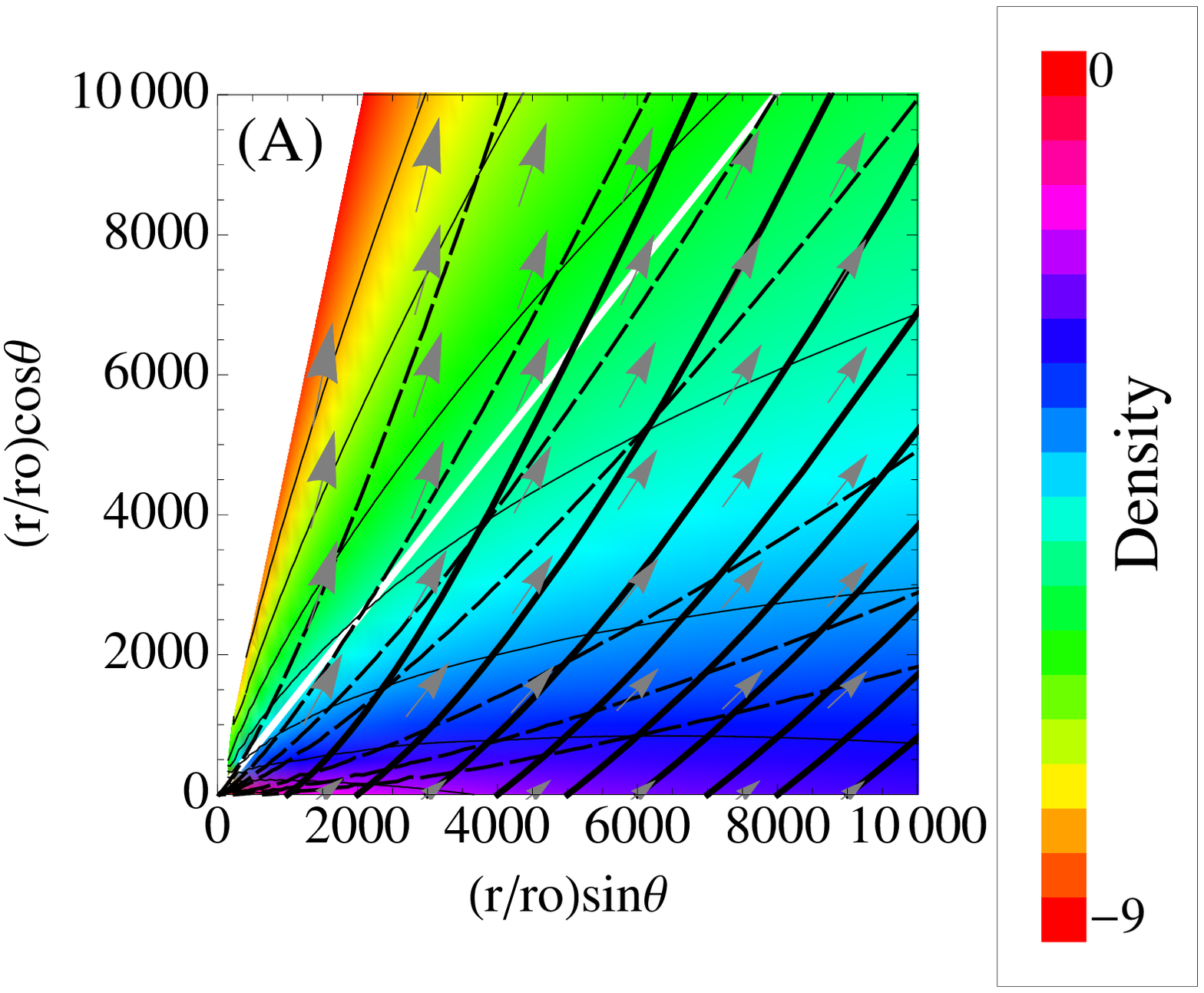} &
\includegraphics[trim=1.9in 0.0in 0in
0in,keepaspectratio=false,width=3.0in,angle=-0,clip=false]{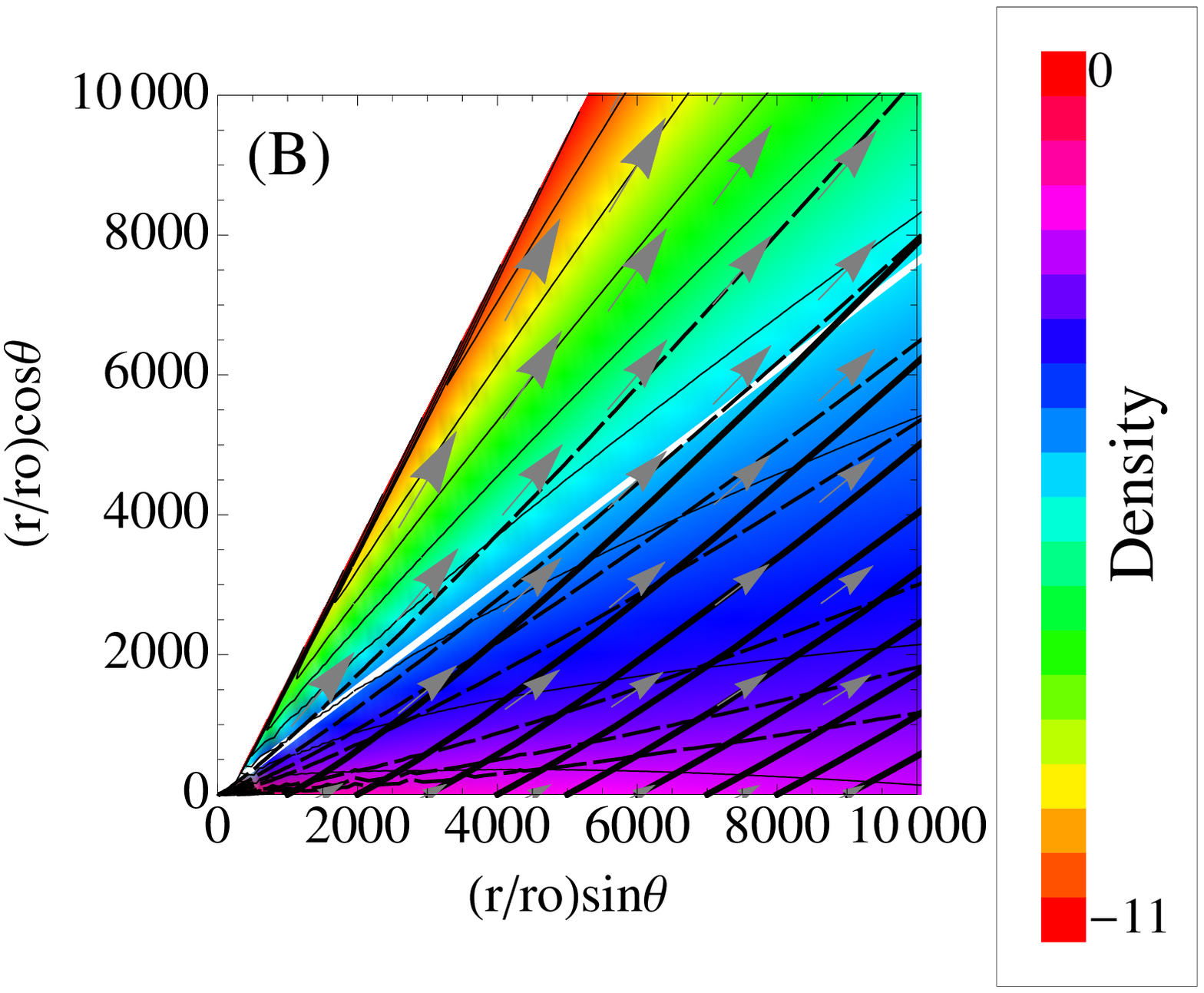} 
\\
\includegraphics[trim=0in 0in 0in
0in,keepaspectratio=false,width=3.9in,angle=-0,clip=false]{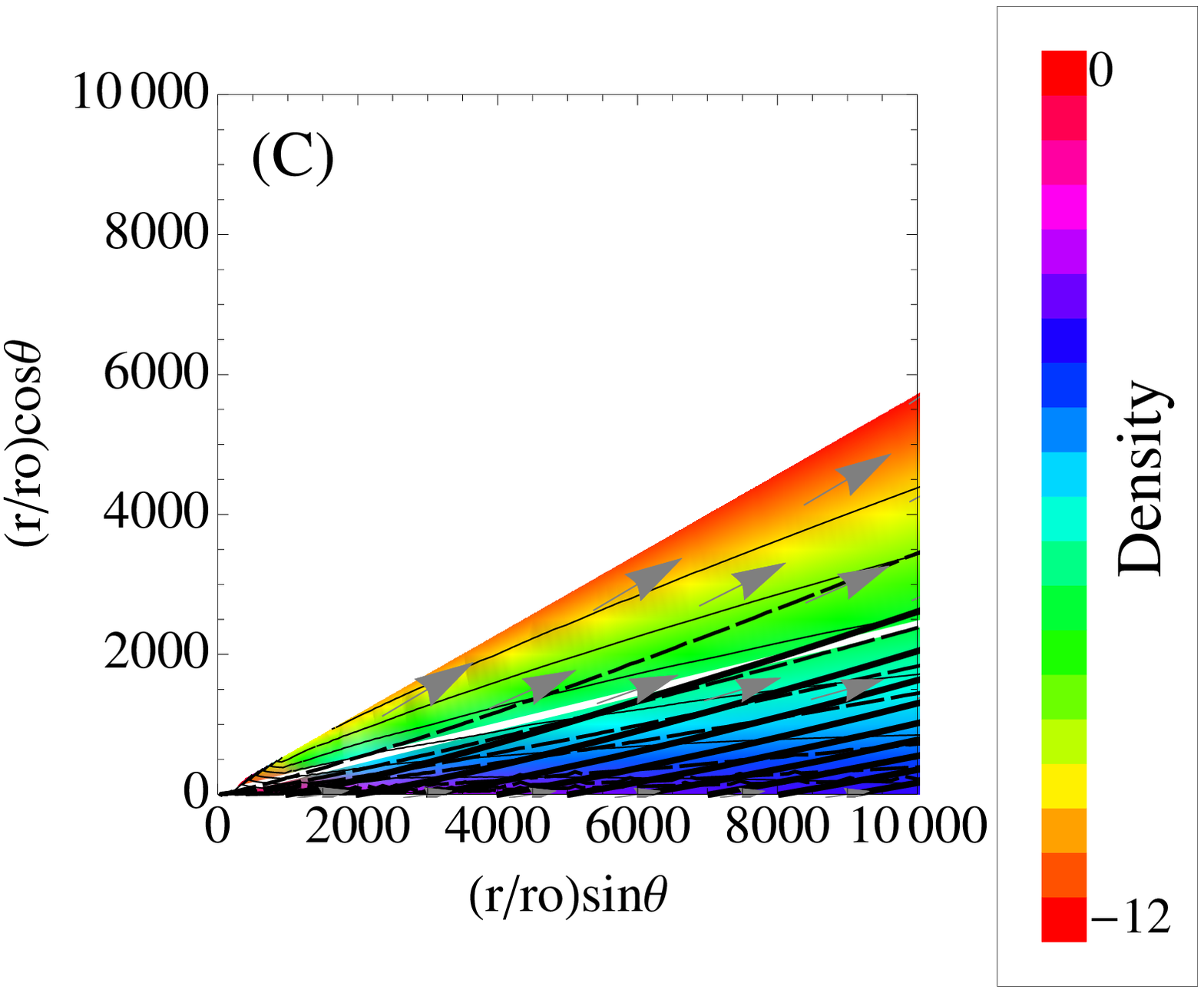} &
\includegraphics[trim=1.9in 0.0in 0in
0in,keepaspectratio=false,width=3.0in,angle=-0,clip=false]{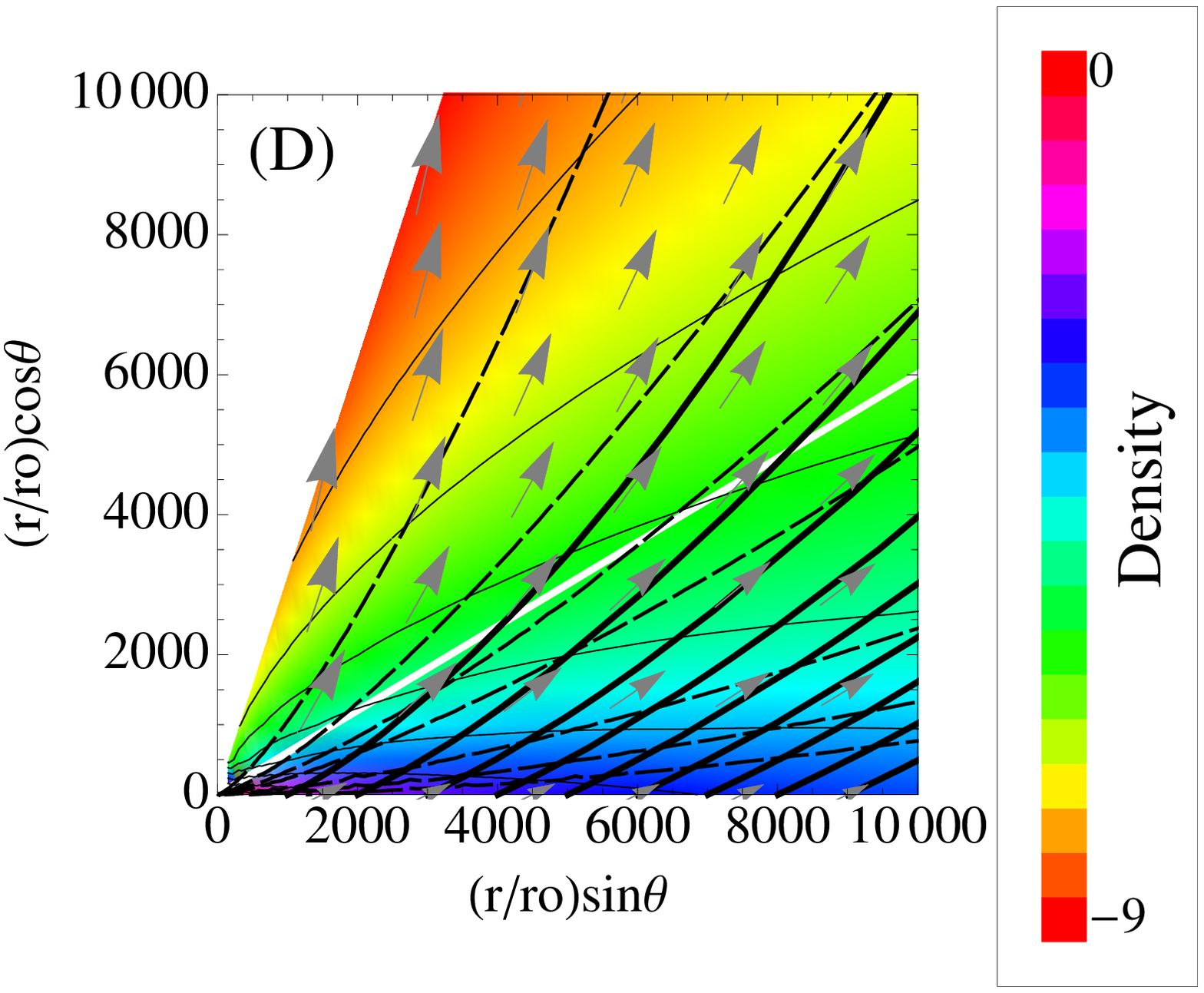} 
\end{array}$
\end{center}
\caption{\baselineskip=11pt Dimensionless quantities of the MHD winds in the poloidal
plane ($r \sin \theta, r \cos \theta$) showing density distribution $\log [m_p
n(r,\theta)/(B_o/v_o)^2]$ along with its contours (solid)
superimposed by wind velocity $\mathvp(r,\theta)/v_o$
(arrows), its contours (dashed) and streamline geometry
$r(\theta)$ (thick solid) for the models (A)-(D). A white line
denotes the \Alfven\ surface. Distance is normalized to $r_o$ around
a BH. The density is normalized to its maximum value. The wind parameters in
each model are listed in Table~\ref{tab:tab1}. } \label{fig:fig1}
\end{figure}

\section{Results}

Employing the framework discussed above, we have studied the
structure and geometry of winds with two distinct density profiles,
namely, winds with $n(r) \propto r^{-1}$ (i.e. $q \sim 1$;
CL94,KK94), because they are favored by the X-ray absorber data, and
winds with $n(r) \propto r^{-3/2}$ (i.e. $q \sim 3/4$; BP82),
because they are best known and most widely cited.
Table~\ref{tab:tab1} shows the functional form of $n(r)$, the values
of  $F_o, H_o, \Omega_o, J_o$, the angle of the \Alfven\ surface
$\theta_{\rm A}$, the wind asymptotic (i.e. at $z/r_o=10^4$)
poloidal velocity normalized to $v_o$, $v_{\rm p,4}/v_o$, and the
wind opening angle $\theta_{\rm open,4}$ at $z/r_o=10^4$
for the wind element launched at the fiducial radius $r_o$.
%
%
The corresponding poloidal  wind structure of all four cases is
shown in Figure~\ref{fig:fig1} in linear scale in the computational
domain $1-10^4 r_o$.
The figure shows the poloidal density distribution
normalized to its maximum value $n_o = B_o^2/(m_p v_o^2)$ (in fact
it shows the value of $\log [m_p n/(B_o/v_o)^2])$ along with its
contours (solid thin lines), the wind streamlines $r(\theta)$ (solid
thick lines) equally spaced by $\Delta r/r_o = 10^3$, the poloidal
velocity vectors $\mathvp/v_o$ (gray arrows) and the velocity
contours (dashed lines). A white solid line denotes the \Alfven\
surface where the plasma poloidal speed $v_p$ reaches the local
\Alfven\ speed $v_A \equiv B_p / \sqrt{4\pi m_p n}$. As mentioned earlier in \S 2 we decided to artificially truncate the super-\Alfven\ wind solution at a critical latitude $\theta_c \equiv \cot^{-1} (100 \cot \theta_A)$ because of uncertainty in self-similarity as well as trans-fast nature of the solutions where $\theta_A$ denotes the latitude corresponding to the \Alfven\ point.

\clearpage

The values of variables in these figures in physical units can be
obtained from equations~(\ref{eq:B}) through (\ref{eq:beta}) with an
appropriate normalization for the plasma density and velocity. While
the velocity normalization can be obtained, independently of the
mass of the accreting object $M$, once a choice of the fiducial
radius $r_o$ (close to the \sw radius $R_S$) is made (where $v_o \simeq c
\,(r_o/R_S)^{-1/2}$), the normalization of the density (and the
ensuing ones of the pressure and magnetic field) requires a value
for the BH mass $M$ and the dimensionless accretion rate $\dot m$
(normalized to its Eddington value). The normalization of the
density can be obtained by considering that for accretion or outflow
at the local escape velocity and at the Eddington rate, the Thomson
optical depth of the flow is equal to one or greater at $r_o \simeq R_S$; this
implies that $n_o \simeq \dot m/\sigma_T R_S$ or $n_o \simeq 5
\times 10^{10} \, ( \dot m / M_8 ) \; {\rm cm}^{-3}$. With these
considerations and assuming magnetic field in equipartition with the
kinetic energy density, i.e., $n m_p v_o^2/2 = B^2/8\pi$, in a flow
with $\dot m \simeq 1$ and $M \simeq 10^8 M_{\odot}$ we obtain the
values
\begin{eqnarray}
n(r,\theta) & \simeq & 4.1 \cdot 10^{10} 
\left(\frac{B_o}{10^4\textrm{G}}\right)^2
\left(\frac{v_o}{0.4c}\right)^{-2} \left(\frac{r}{r_o}\right)^{2q-3}
\left(\frac{\tilde{n}(\theta)}{\tilde{n}(90^{\deg})}\right) ~~\textrm{cm$^{-3}$} \ ,\\
& \simeq & 5 \times 10^{10} \, \left(\frac{\dot m}{M_8} \right)
\left(\frac{r}{r_o}\right)^{2q-3}
\left(\frac{\tilde{n}(\theta)}{\tilde{n}(90^{\deg})}\right)
~~\textrm{cm$^{-3}$}\label{eq:n2} \\
p(r,\theta) & \simeq & 2.1 \times 10^{4} 
\left(\frac{K}{0.01}\right)
\left(\frac{B_o}{10^4\textrm{G}}\right)^2
\left(\frac{r}{r_o}\right)^{2q-4}
\left(\frac{\tilde{n}(\theta)}{\tilde{n}(90^{\deg})}\right)^\Gamma
~~\textrm{dyne~cm$^{-2}$} \ , \label{eq:p2} 
\end{eqnarray}
%
%
with the angular parts of $\tilde{n}, \tilde{p}$ and $\tilde{B}$ set
equal to one on the disk surface $(\theta = 90\degr$) and their
$\theta-$dependence produced by the solution of the MHD equations.
The normalization of the quantities given above, appropriate for $M
= 10^8 M_{\odot}$, scales like $1/M$, i.e. inversely with the black
hole mass, while the plasma $\beta$ is independent of it.

\begin{deluxetable}{ccccccccccc}
\tabletypesize{\scriptsize} \tablecaption{Characteristics of baseline MHD
wind models. \label{tab:tab1}} \tablewidth{0pt} \tablehead{Model &
$n(r)$ & $F_o$ & $H_o$ & $\theta_o$ & $\Omega_o$ & $J_o$ & $\theta_A$
 & $v_{p,4}/v_o$ & $\theta_{\rm open,4}$}
\startdata
A & $r^{-1}$ & 0.065 & -1.7 & 1.276 & 1.016 & -1.51 & $38\degr$  & $4$ & $7\degr$   \\
B & $r^{-1}$ & 0.03 & -2.455 & 1.872 & 1.024 & -1.52 & $53\degr$  & $7$  & $28\degr$   \\
C & $r^{-1.5}$ & 0.05 & -3.152 & 5.7 & 1.032 & -1.53 & $76\degr$  & $6$ & $47\degr$  \\
D & $r^{-1.5}$ & 0.1 & -1.68 & 2.105 & 1.015 & -1.515 & $59\degr$  & $3$ & $6\degr$
\\
\enddata
\end{deluxetable}

The properties of the models of Figure \ref{fig:fig1} are summarized
in Table~1 (see the Appendix for a brief comparison with BP82
wind parameters).  All these models have approximately the same
values of $\Omega_o$ and $J_o$; models (A) and (B) have density
profile $n(r) \propto r^{-1}$ (CL94; KK94), while for models (C) and
(D) $n(r) \propto r^{-3/2}$ (BP82). The larger value of $H_o$ in
model (B) results in less collimation, i.e. larger asymptotic
opening angle ($\theta_{\rm open,4} \sim 28\deg$), compared with
that of model (A) ($\theta_{\rm open,4} \sim 7\deg$) which has a
smaller $H_o$ value. For the same reason the \Alfven\ surface of (A)
lies at smaller value of $\theta$ ($\theta_A \sim 38 \degr$) than of
(B) ($\theta_A \sim 53 \degr$). Finally, the combination of larger
$H_o$ and smaller mass loading leads to larger asymptotic value in
(B) ($v_{\rm p,4}/v_o \sim 7$) compared to that of (A) ($v_{\rm
p,4}/v_o \sim 4$). The models with $n(r) \propto r^{-3/2}$ (C,D)
have relative dependencies similar to those of (A,B) and are
summarized in Table 1. Finally, for models with similar values of
$\Omega_o, J_o$ and $H_o$ but different density profile, namely (A)
and (D), while the asymptotic opening angles at $z = 10^4 r_o$ are
similar, their \Alfven\ angle is smaller in (A) than (D) presumably
as the result of the different poloidal current distribution of
these two solutions
via a coupling effect between $F_o$ and $H_o$.

\begin{figure}[ht]
\begin{center}$
\begin{array}{cc}
\includegraphics[trim=0in 0in 0in
0in,keepaspectratio=false,width=2.9in,angle=-0,clip=false]{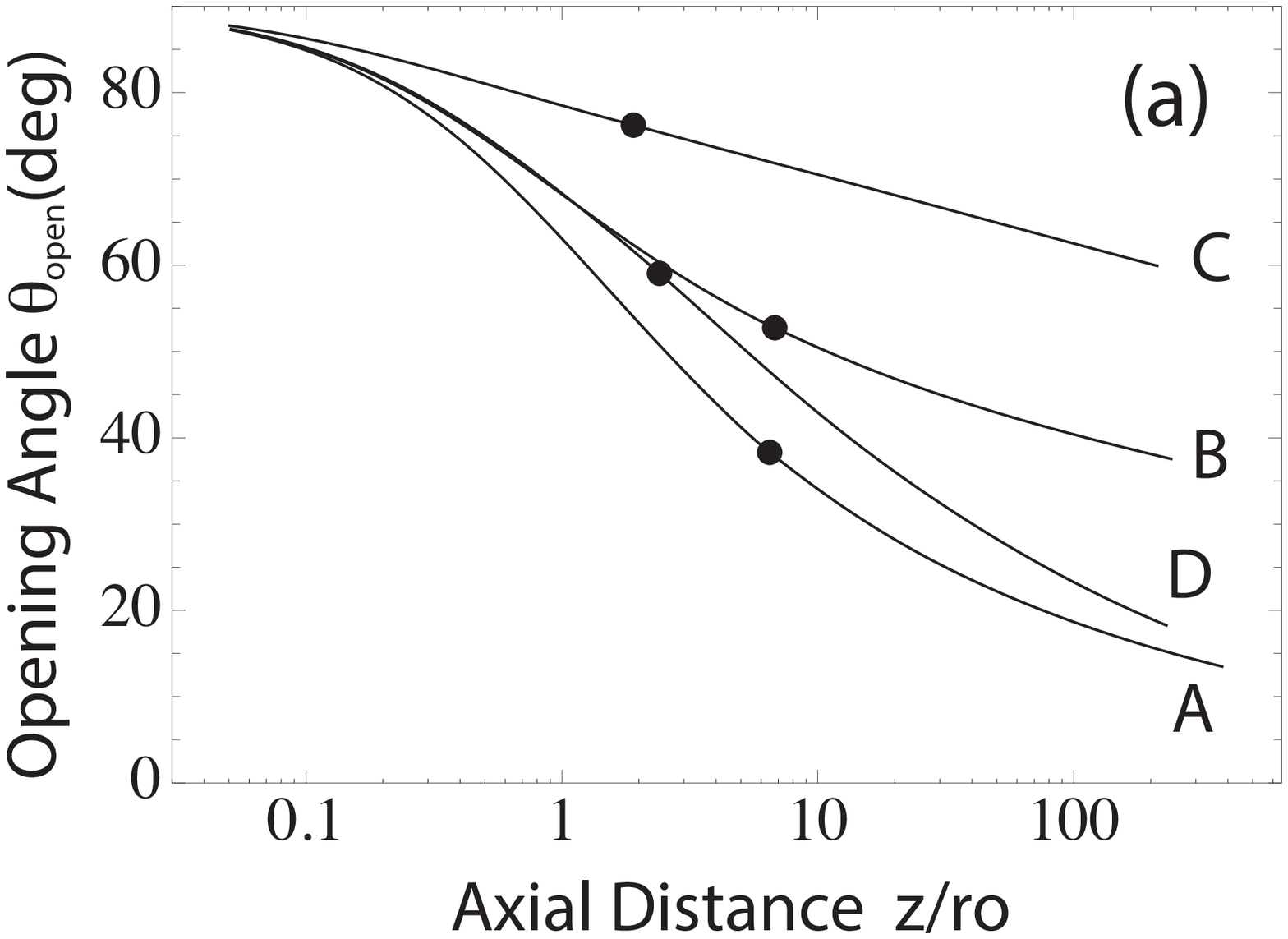} 
&
\includegraphics[trim=0in 0in 0in
0in,keepaspectratio=false,width=2.9in,angle=-0,clip=false]{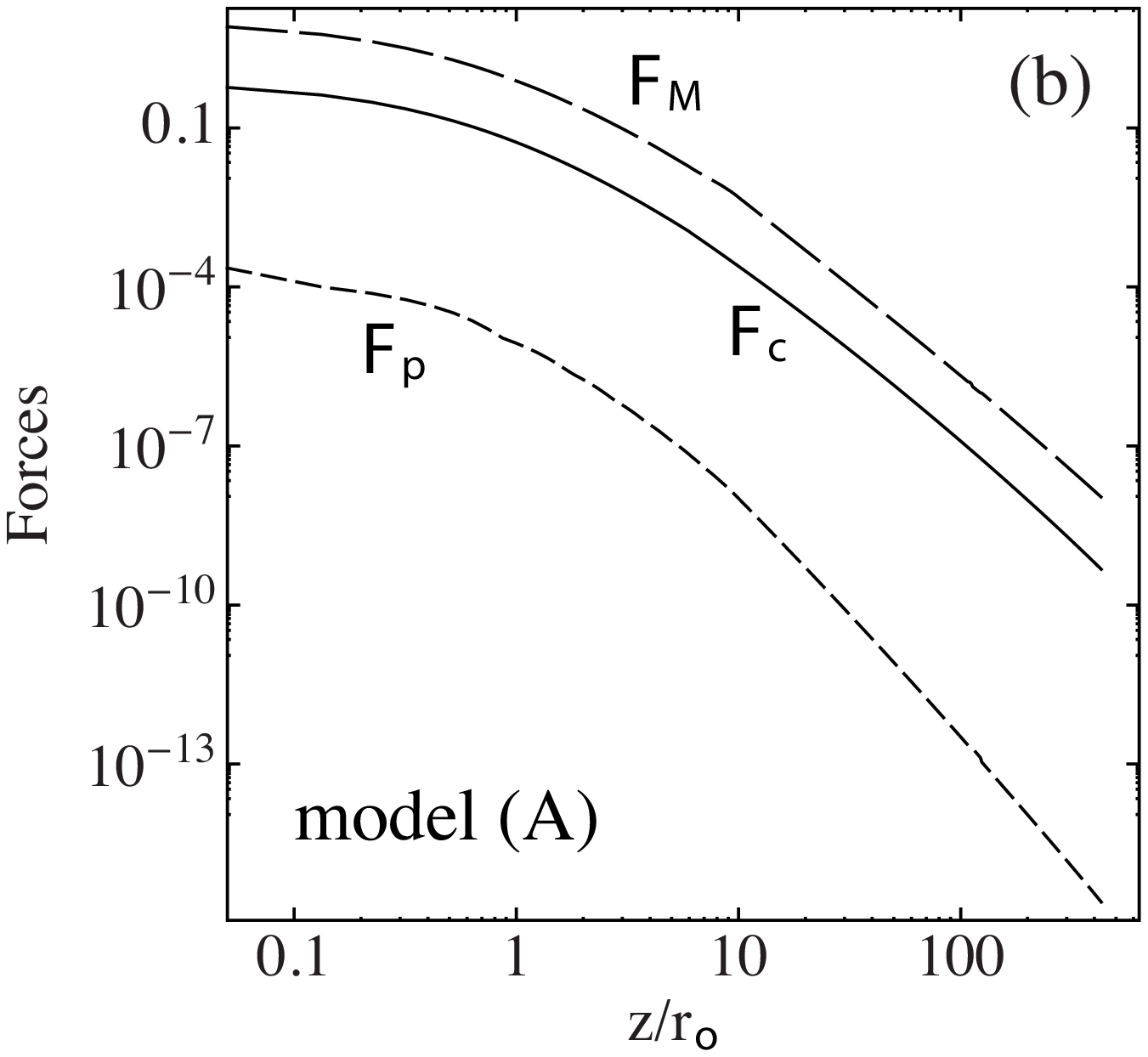} 
\end{array}$
\end{center}
\caption{\baselineskip = 11pt (a) Disk-wind opening angle $\theta_{\rm open}$ as a function of
axial distance $z$ (up to $\theta_{c} $) along the innermost streamline (with
$\Psi=\textrm{const}$) of winds launched at $r=r_o$ for
models (A)-(D). (b) Force components tangent to the poloidal field line for
model (A) showing $\mathJ \times \mathB$ term $F_M$ (dashed), centrifugal term $F_c$ (solid) and gas pressure term $F_p$ (dotted). The wind
parameters are listed in Table~\ref{tab:tab1}. } \label{fig:fig2}
\end{figure}

These results are summarized in Figure~\ref{fig:fig2}a which shows
the outflow opening angle $\theta_{\rm open}$ along the innermost
streamline, originating at $r=r_o$ (these curves correspond to the
innermost streamlines  in Figure \ref{fig:fig1}) as a function of
axial distance $z/r_o$ for models (A)-(D) with dots denoting the
\Alfven\ points. The effects of the higher $H_o$ in
models (B), (C) is apparent there. 
%

Our interest of driving mechanism focuses primarily on $J \times B$ force  while not discussing the centrifugal force in details. 
In the wind model considered here, toroidal wind velocity scales as $v_\phi \propto r^{-1/2}$ along a given LoS. Therefore, centrifugal force term goes as $\sim \rho v_{\phi}^2/r \propto r^{-3}$ along the LoS and magnetic force term also goes as $\nabla B^2 \propto r^{-3}$ for $n \propto r^{-1}$. 
The relative strength of centrifugal term can be dominant at smaller radii at least within the \Alfven\ surface. 
This behavior is indeed confirmed in Figure~\ref{fig:fig2}b comparing the force components tangent to the poloidal field line (i.e. accelerating components) following \citet{LR13}. The relative strength of each force component and its profile is consistent with those in other work (e.g. Fig.~9 in Porth \& Fendt~2010) in that the magnetic part can be the dominant component even at smaller radii. \citet{Lii12} also argues in their 2.5D MHD jet simulations that the centrifugal force is cancelled by gravity thus the jet is driven by a purely magnetic force. This mechanism is similar to the inner disc wind model discussed in \citet{Lovelace91} and observed in simulations of conical winds in \citet{Romanova09}.

\begin{figure}[ht]
\begin{center}$
\begin{array}{cc}
\includegraphics[trim=0in 0in 0in
0in,keepaspectratio=false,width=2.8in,angle=-0,clip=false]{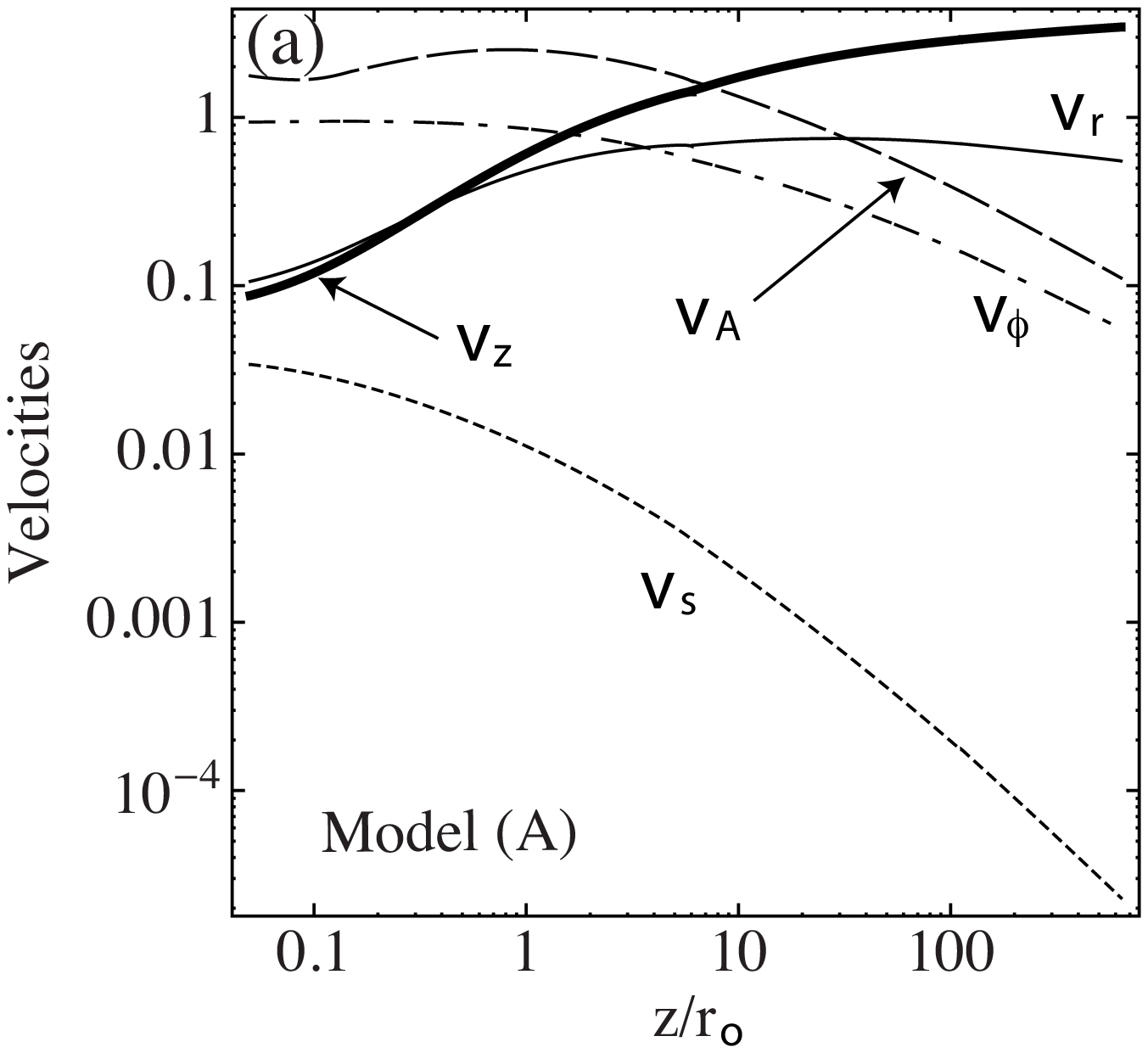} 
& \includegraphics[trim=0in 0in 0in 0in,keepaspectratio=false,width=2.8in,angle=-0,clip=false]{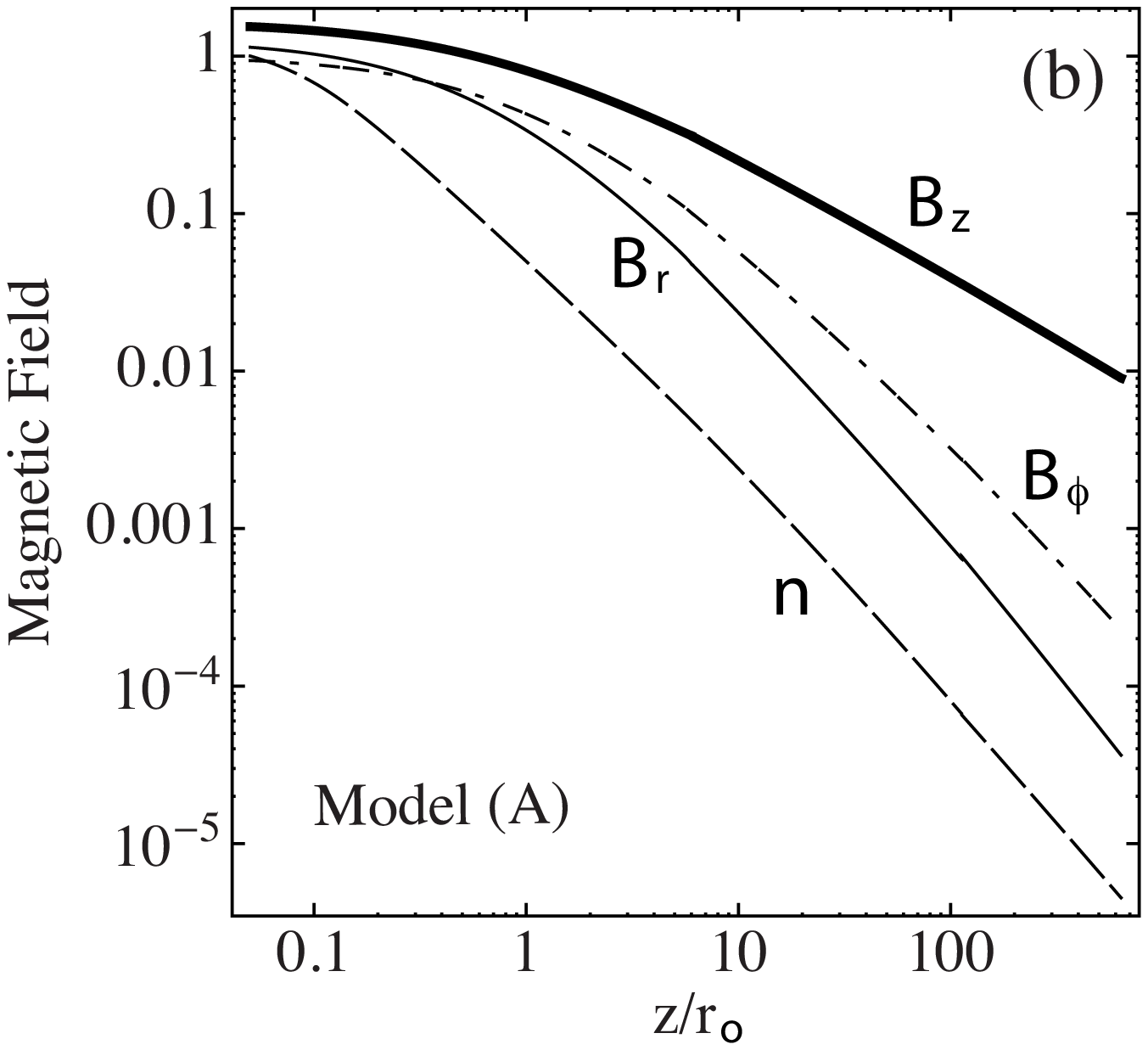} 
\\
\includegraphics[trim=0in 0in 0in
0in,keepaspectratio=false,width=2.8in,angle=-0,clip=false]{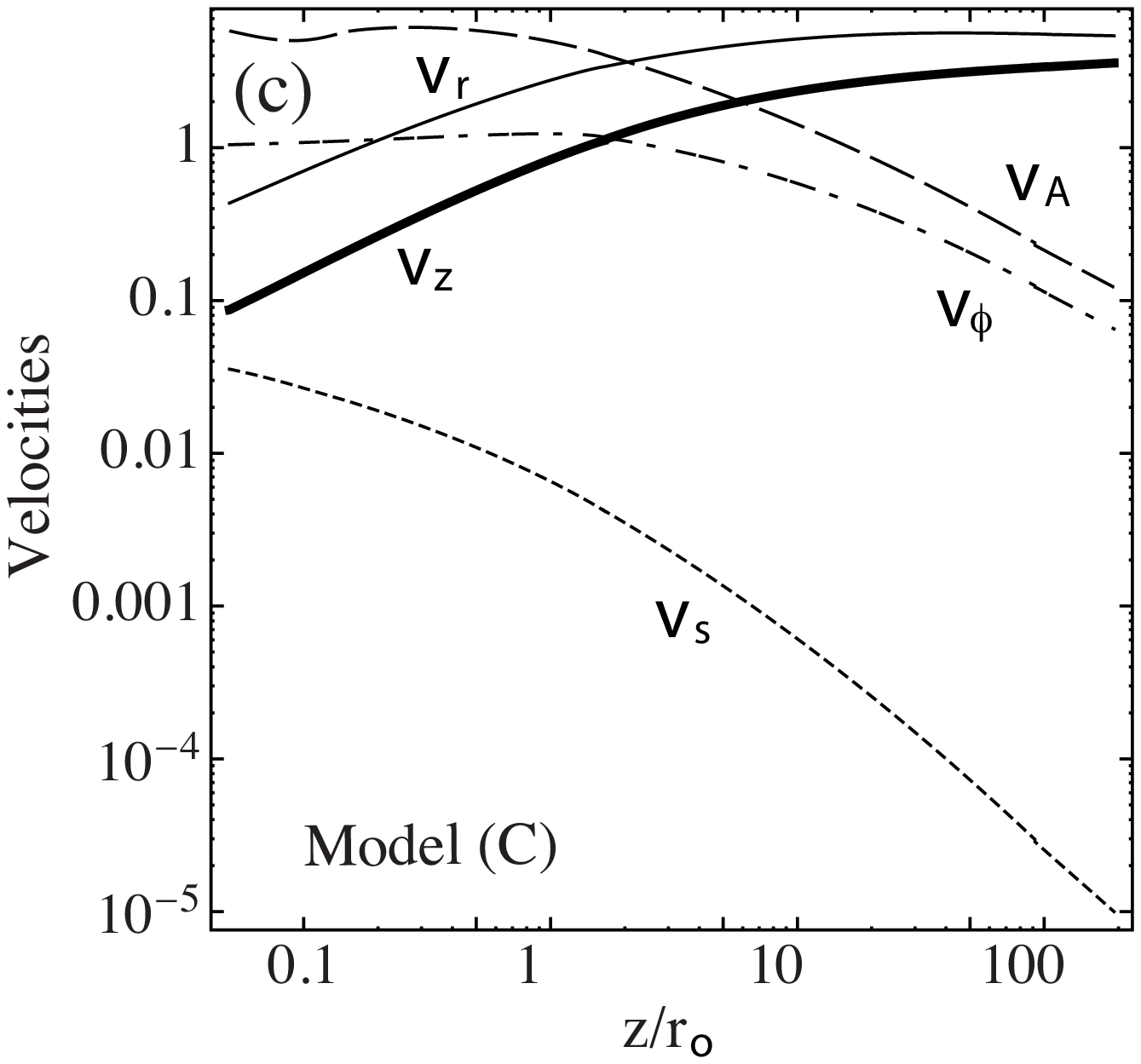} 
& \includegraphics[trim=0in 0in 0in
0in,keepaspectratio=false,width=2.8in,angle=-0,clip=false]{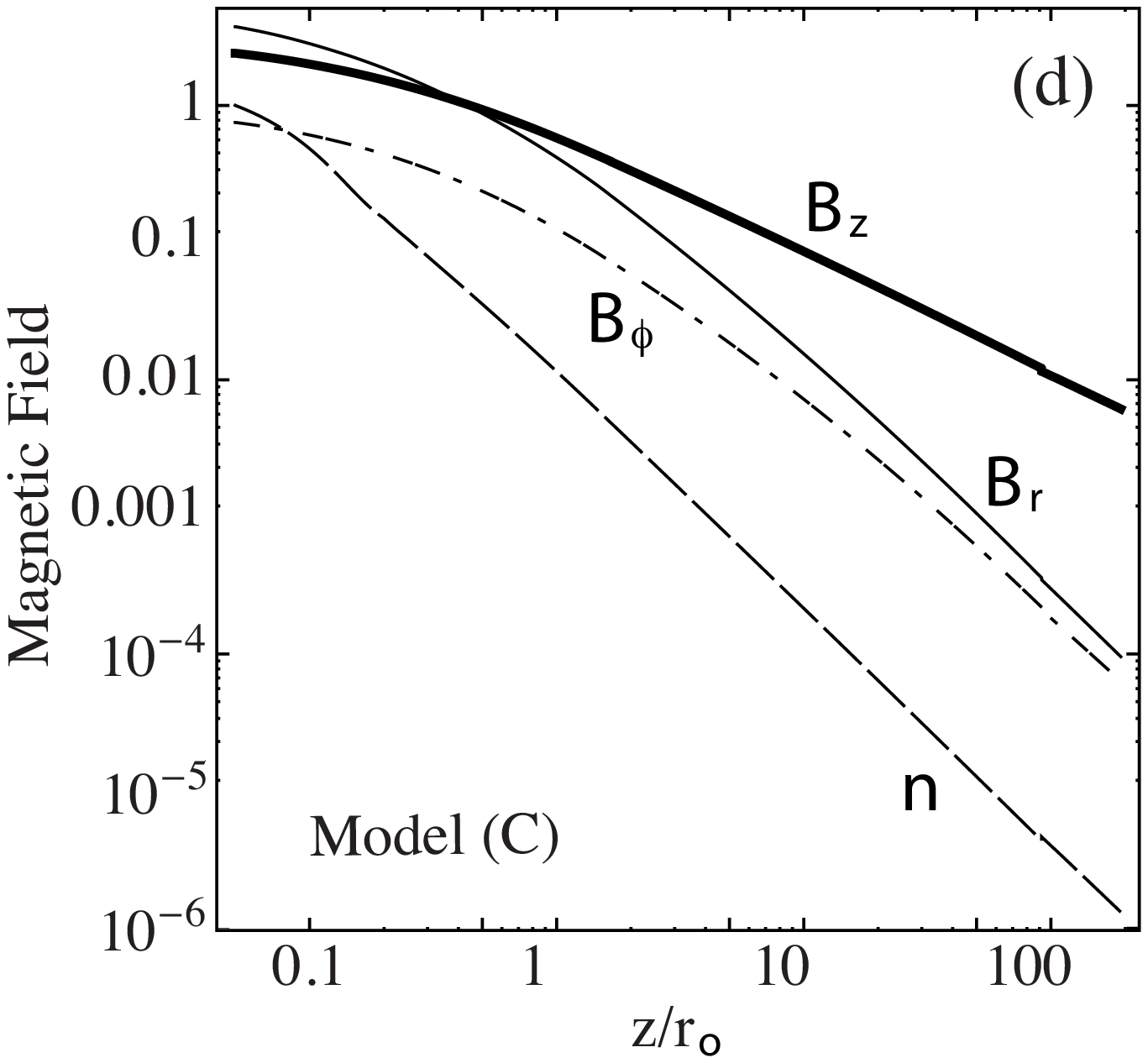} 
\end{array}$
\end{center}
\caption{\baselineskip = 11pt  Fiducial wind velocity of radial $v_r$, toroidal $v_\phi$, axial $v_z$ components along with the poloidal \Alfven\ speed $v_A$ and slow magnetosonic speed $v_s$ normalized to $v_o$ (along the innermost streamline) and the corresponding magnetic field of radial $B_r$, toroidal $B_\phi$ and axial $B_z$ components normalized to $B_o$ as well as density profile $n$ as a function of axial distance $z/r_o$. We show for model (A) in the upper panels while (C) in the lower panels. The wind
parameters are listed in Table~\ref{tab:tab1}. } \label{fig:fig3}
\end{figure}

\clearpage

Focusing more on the global properties of magnetized wind, Figure~\ref{fig:fig3} shows  kinematic profiles $\mathv$ of the wind and magnetic field $\mathB$ as well as wind density profile $n$ as a function of axial distance $z/r_o$ for model (A) in the upper panels and (B) in the lower panels.
%
%
Due to a geometrical difference between the two models, radial wind velocity still exceeds the axial one in model (C). While velocity profiles $\mathv$ are independent of the models since we assume Keplerian boundary condition at the disk surface, magnetic  field profiles $\mathB$ and density $n$ are indeed dropping faster (i.e. $\propto z^{-3/2}$) in model (C) as expected.   
It is seen in both models that the wind undergoes a rapid acceleration phase from
the base through the \Alfven\ point (i.e. the intersection between $v_z$ and $v_A$), after which the wind
approaches asymptotically a coasting speed at $v_{\rm p,4}/v_o \sim
4$ in model (A). The predominant initial orbital motion ($v_\phi$) of the wind at the time of launch is found to be efficiently converted into axial motion ($v_z$). These profiles follow $|\mathv| \propto z^{-1/2}, |\mathB| \propto z^{-1}$ and $n \propto z^{1/2}$  at large distances in  consistence with the chosen $1/r$ self-similarity in model (A). 

%
%

\begin{figure}[th]
\begin{center}$
\begin{array}{cc}
\includegraphics[trim=0in 0in 0in
0in,keepaspectratio=false,width=3.9in,angle=-0,clip=false]{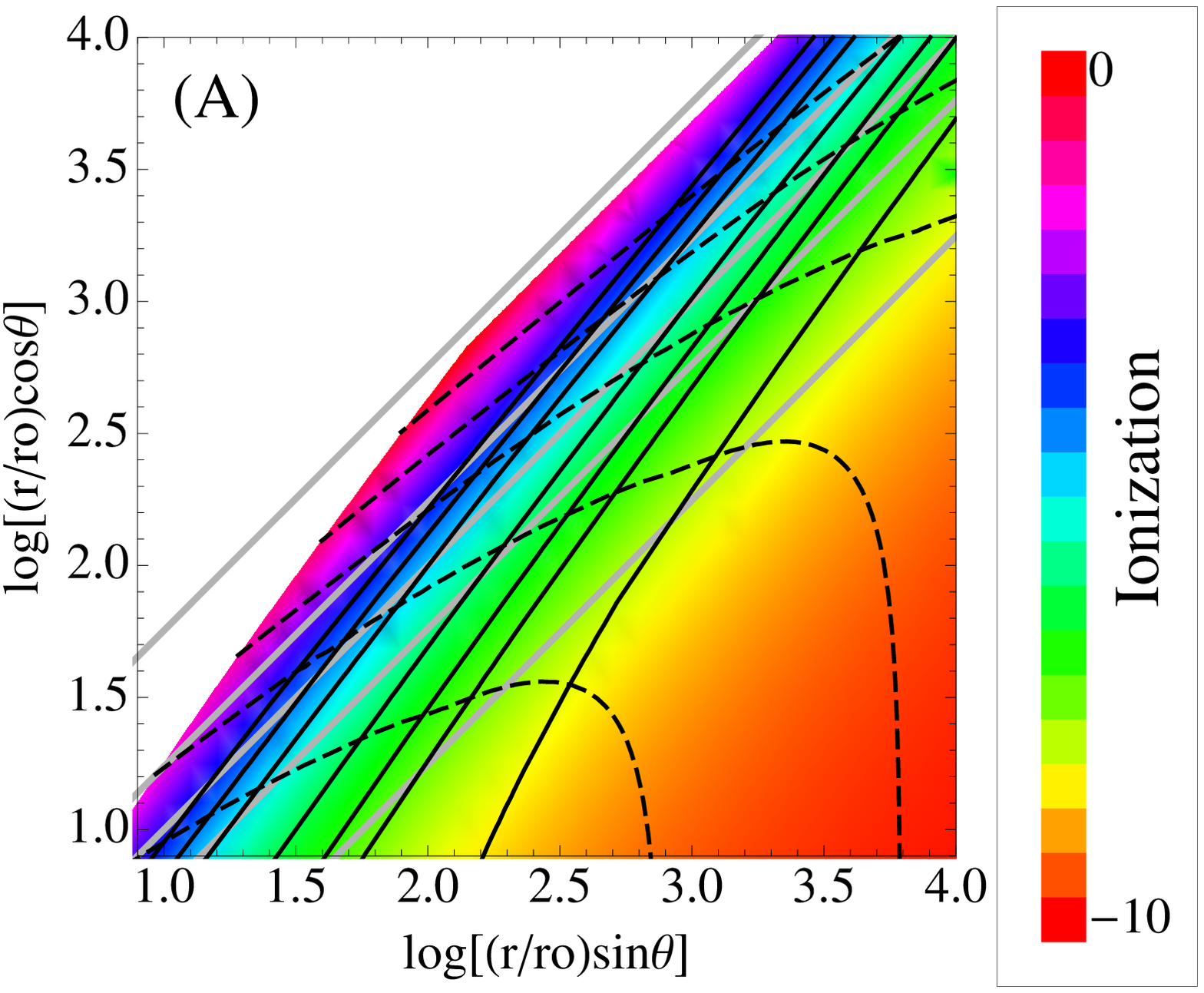} &
\includegraphics[trim=1.9in 0.0in 0in
0in,keepaspectratio=false,width=3.0in,angle=-0,clip=false]{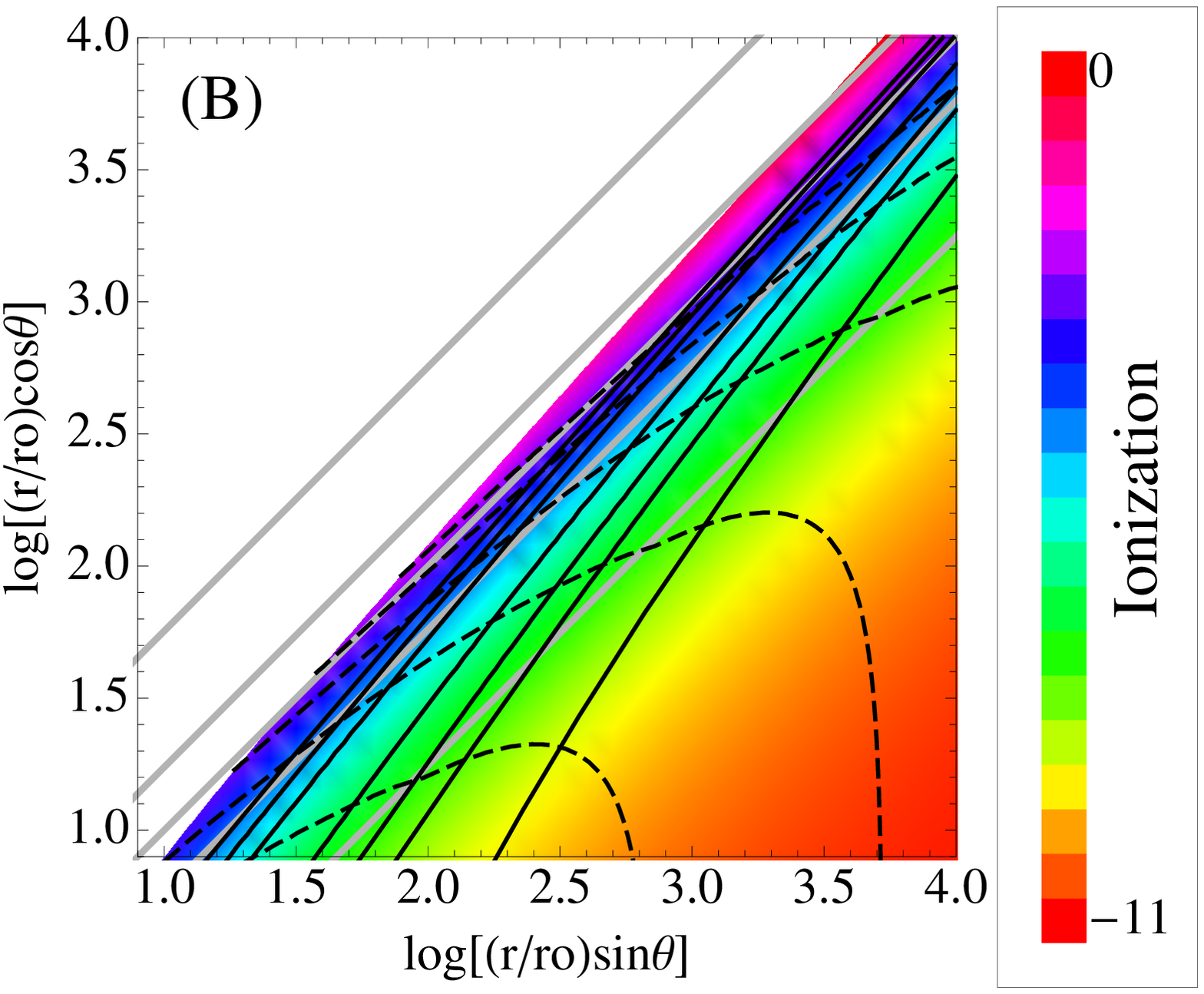} 
\\
\includegraphics[trim=0in 0in 0in
0in,keepaspectratio=false,width=3.9in,angle=-0,clip=false]{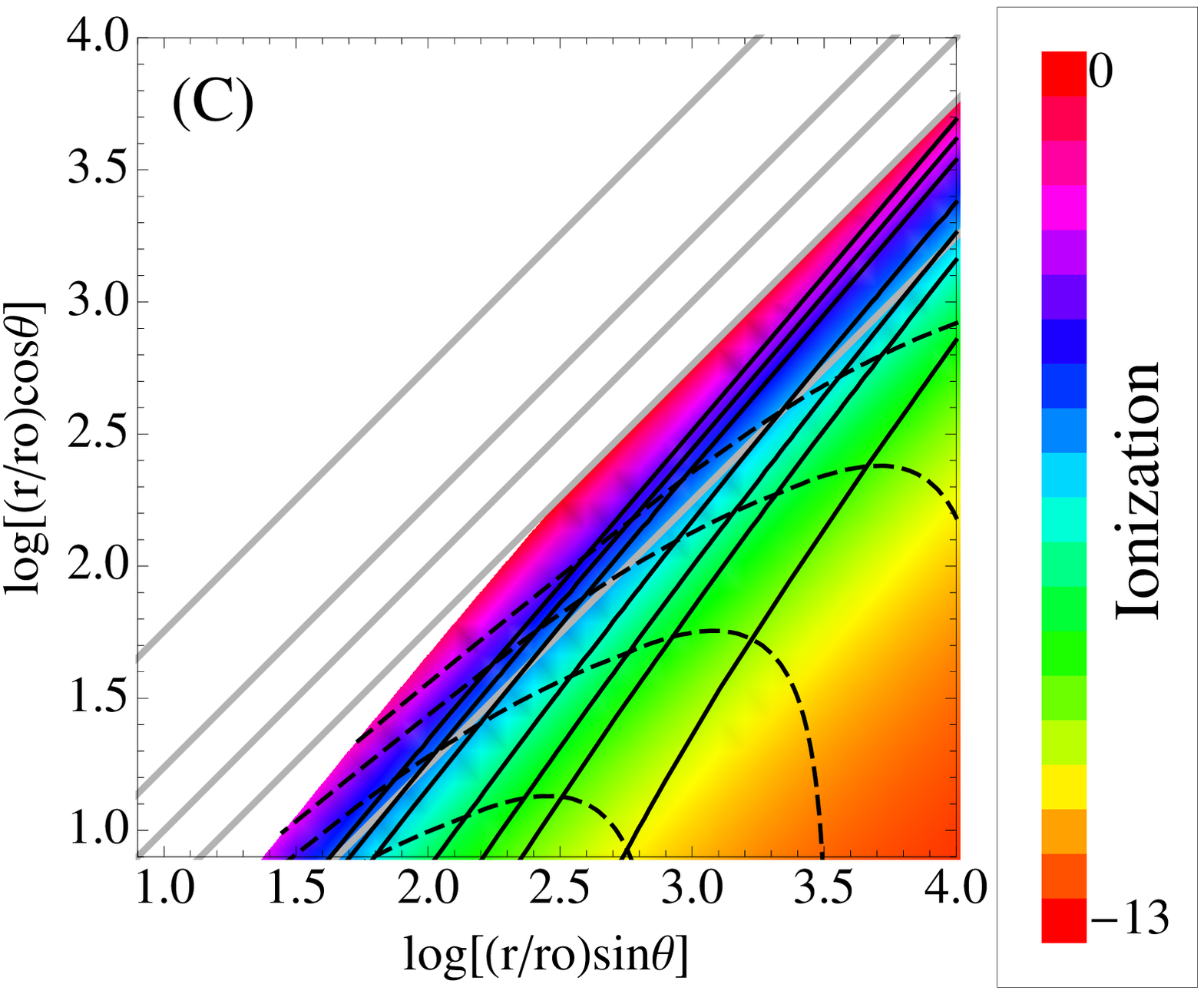} &
\includegraphics[trim=1.9in 0.0in 0in
0in,keepaspectratio=false,width=3.0in,angle=-0,clip=false]{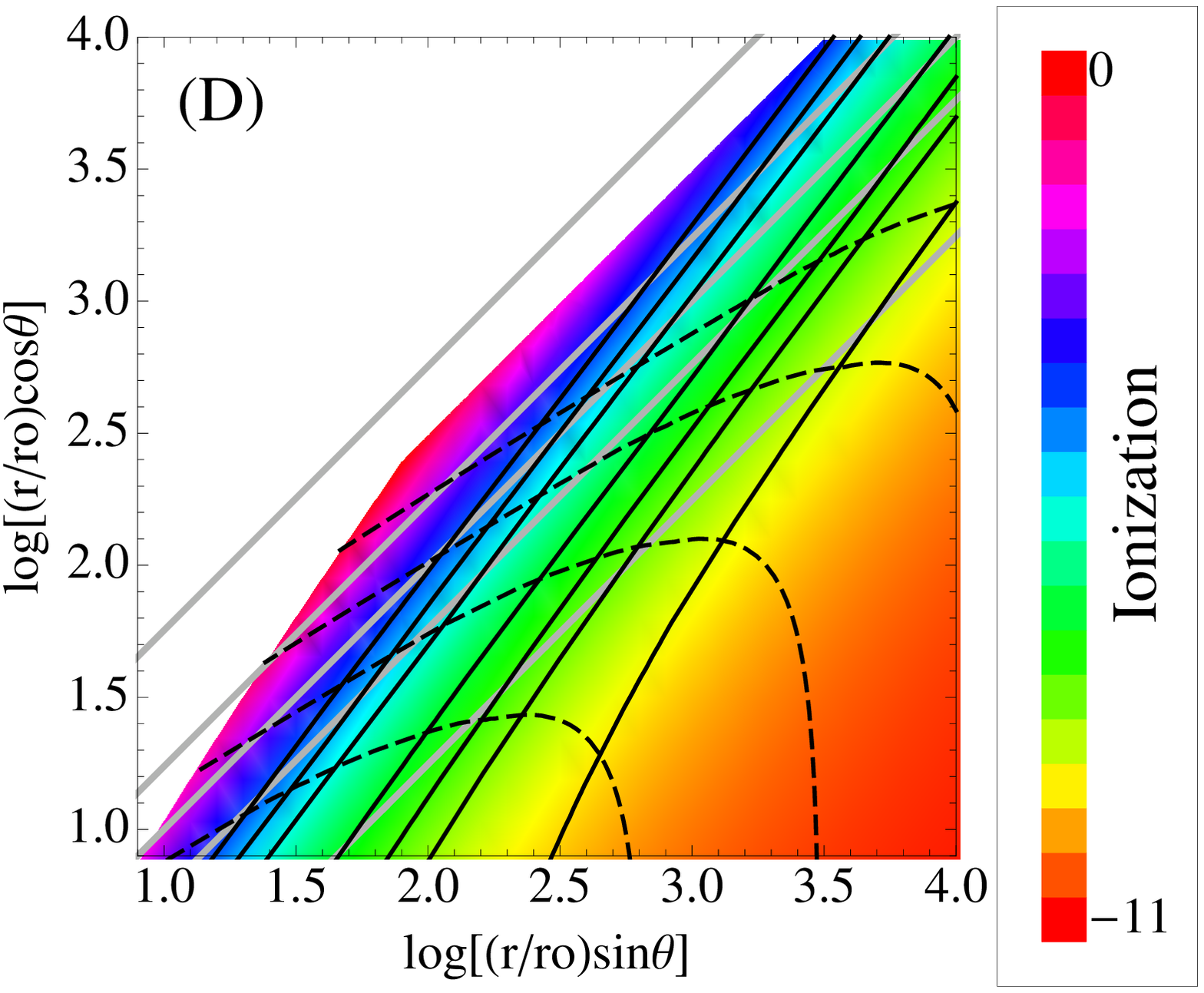} 
\end{array}$
\end{center}
\caption{\baselineskip=13pt Contour plot (in log-log space) of models (A)-(D)
corresponding to Figure~1 showing
LoS poloidal velocity (solid) of $v/c = 0.5, 0.3, 0.1, 0.05, 0.03, 0.01$
(decreasing outward) and the density
(dashed) $\log n = -1, -2, -3, -4, -5, -6, -7$ (decreasing upward)
superimposed by color-coded ionization parameter $\log \xi$ which
is normalized to their maximum value for $\dot{m}=0.1$.
Five LoS angles are denoted by thick gray lines ($\theta=10\degr,
30\degr, 45\degr, 60\degr, 80\degr$ from top to bottom). The wind
parameters are listed in Table~\ref{tab:tab1}.    } \label{fig:fig4}
\end{figure}


\clearpage

As discussed in \S 1, the observations and physics of WAs provide
mainly their ionization parameter $\xi$ and hydrogen equivalent
column density $N_H$, and to the extent that can be measured, also
the plasma velocity along the observer's LoS. Therefore, in order to
relate the above models to observations, we plot in Figure
\ref{fig:fig4} (in log-log space) the structure of ionization
parameter $\log \xi$[erg~cm~s$^{-1}$] for $\dot{m}=0.1$, along with
the normalized density contours (dotted curves for $10^{-1},
10^{-2}, 10^{-3}, 10^{-4}, 10^{-5}, 10^{-6}, 10^{-7}$ from bottom to
top) and velocity contours (solid black curves for $v/c = 0.5, 0.3,
0.1, 0.05, 0.03, 0.01$ from innermost to outermost) of the same
winds shown in Figure~\ref{fig:fig1}.
Here, we only consider geometrical dilution factor (see footnote 1) whereas in our photoionization calculations the opacity of the wind material was included (FKCB10a, FKCB10b). 
Because these quantities span a range of several decades, it is only
reasonable to present the wind structure in logarithmic coordinates.
Thus color-coded $\log \xi$ is shown in the range $\sim 0$ and
$\sim 10$. To make contact with observations we also draw the
observer LoS of different inclinations (diagonal gray lines
corresponding to $\theta=10\degr, 30\degr, 45\degr, 60\degr,
80\degr$ from top to bottom). One can now see that low inclination
lines of sight intercept mainly high ionization and low column
plasma (the iso-density curves are more closely spaced at the lower
inclination directions). It is also apparent that the wind is highly
ionized near its innermost edge where the velocities are also higher
but the column intercepted is lower. It is not clear whether these
sections of the wind can be detected in actual observations. With
increasing inclination the LoS samples lower values of $\xi$ and
velocity $v$ and higher values of the column $N_H$. This model
therefore associates naturally the observed WAs with objects of
sufficiently high inclination angles.


The two bottom panels of Figure \ref{fig:fig4} correspond to winds
with $n(r) \propto r^{-3/2}$ (models C, D). These panels look quite
similar to those of models (A) and (B) above them. The obvious difference
is the more closely spaced iso-density contours due to the steeper
density profile; as a result most of the column along a given LoS is
concentrated near $\log (r) = 0$. For the cases of both density
profiles,  we see that the higher angular momenta solutions leave a
large fraction of the LoS not intercepting any plasma, at least
within the computational domain. This fact would be of
interest for the statistics of absorbers.

The column of these figures along a given LoS, one of the
fundamental observables of the AGN absorption features, depends, as
discussed above, on the value of (the normalized) mass flux and, as
is apparent in Figure \ref{fig:fig4}, on the winds' poloidal
structure. Clearly models (B) and (C) require observer inclinations
greater than $45\degr$ to produce any significant absorption column.
For model (A), a column of  $N_H \simeq 10^{23} \; {\rm cm}^{-2}$
and velocity $\sim 10,000 - 15,000$ km/s, consistent with the
observations of UFOs, assuming mass flux $\dot m \simeq 1$, demands
an inclination angle $\theta \simeq 30\degr-45\degr$, based on the
density function $n(\theta)$ of FKCB10a, while the observed velocity
implies $r/r_o \simeq (v/c)^{-2} \simeq 10^3$. In model (A) we see
that in this region of parameter space, $\log \xi \simeq 3-4$,
consistent with the presence of an \fexxv\ UFO. 
%
%
Clearly models (B) and (C) have very little column for
inclinations $\theta \lsim 30\degr$. These figures also allow one to
estimate the the velocity of a given ion (i.e. a given value of
$\xi$) at a specific LoS and as such they are of value in setting a
particular set of observations within the framework of these winds.

\begin{figure}[ht]
\begin{center}$
\begin{array}{c}
\includegraphics[trim=-0in 0.0in 0in
0in,keepaspectratio=false,width=4.3in,angle=-0,clip=false]{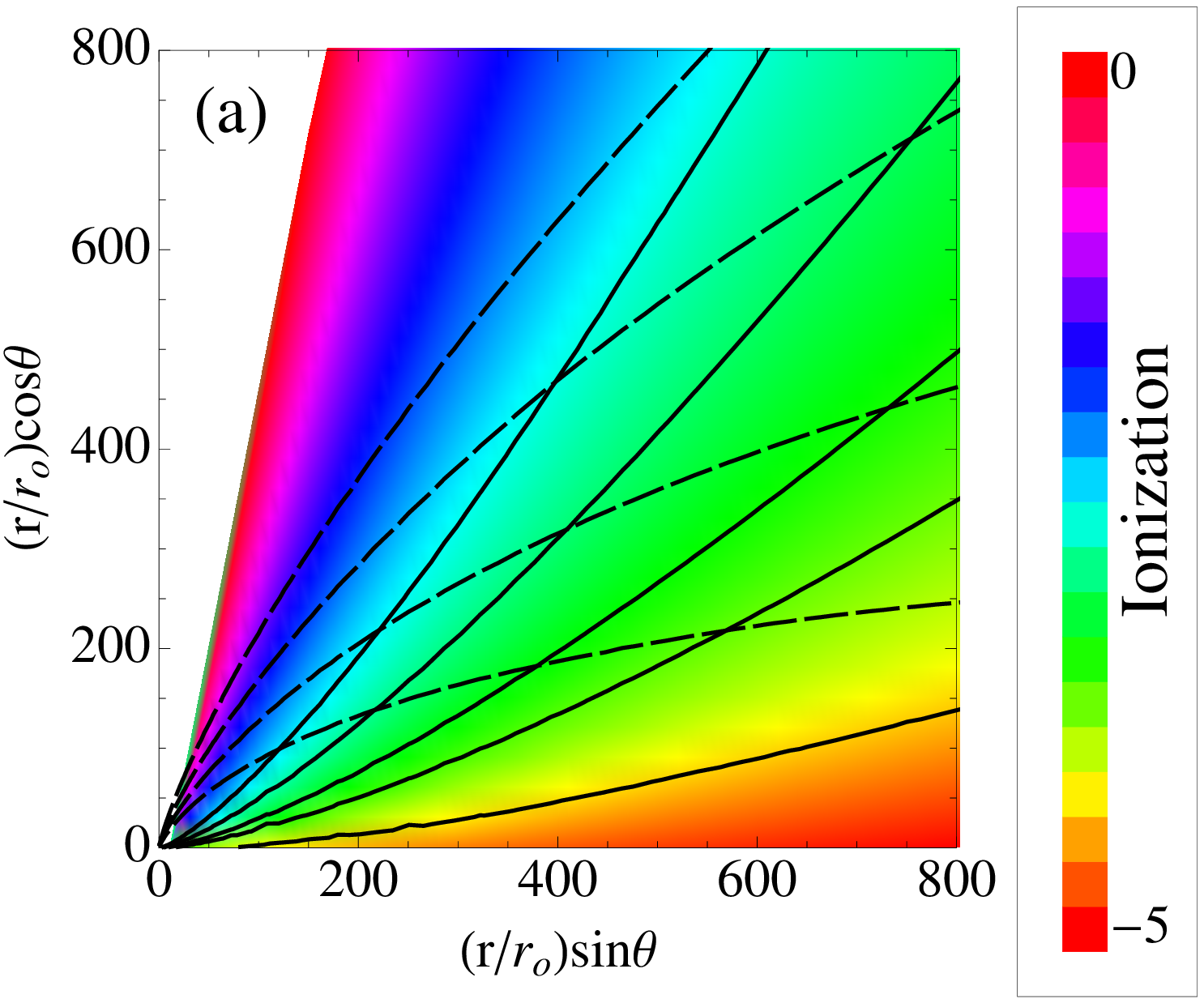}
\includegraphics[trim=2in 0in 0in
0in,keepaspectratio=false,width=1.8in,angle=-0,clip=false]{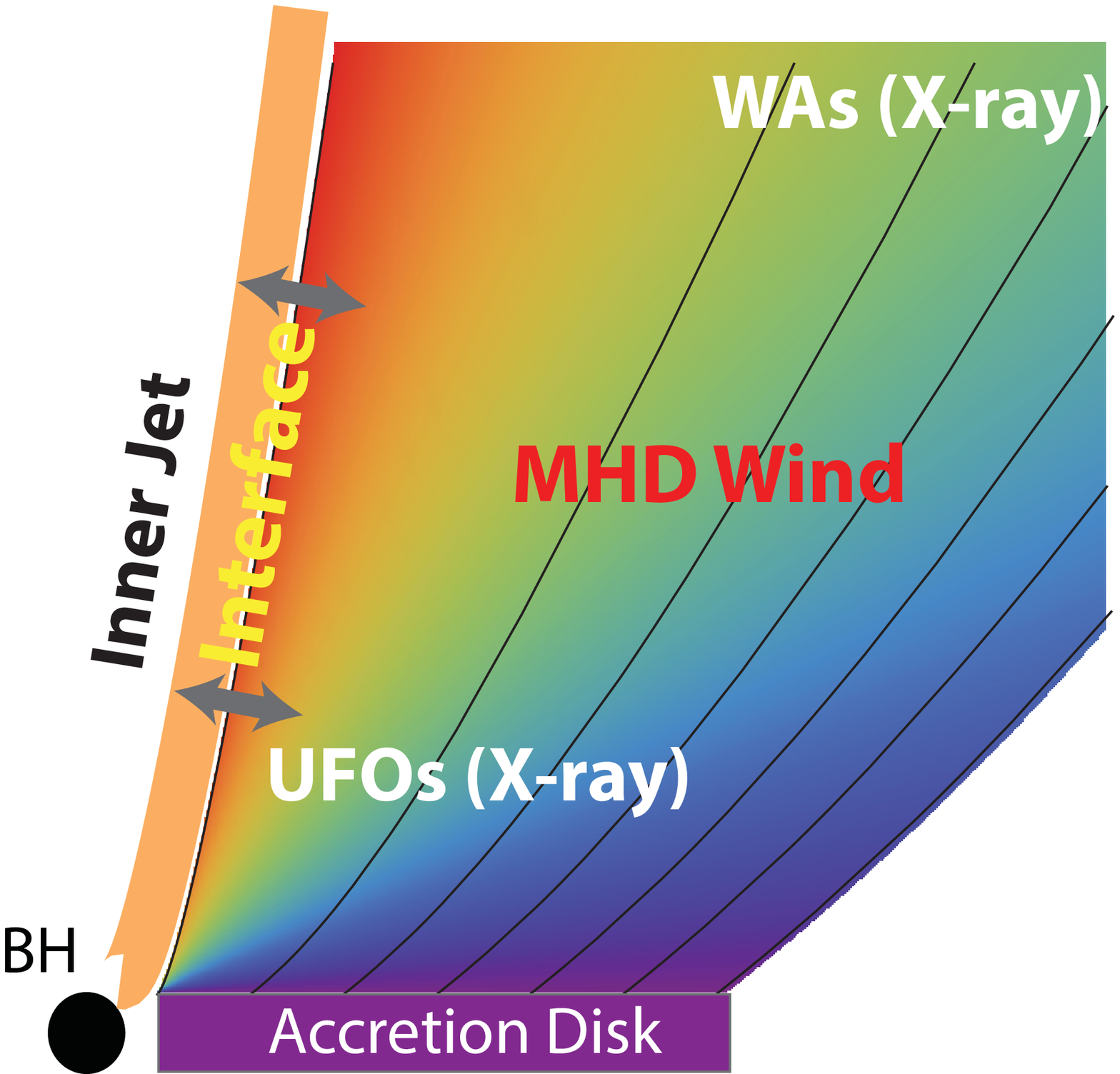}
\end{array}$
\end{center}
\caption{(a) Contour plot of model (A) showing LoS poloidal velocity (from top to bottom in solid)
of $v/c = 0.2, 0.1, 0.05, 0.03, 0.01$ and wind pressure
(from top to bottom in dashed) of $p_{\rm UFO}=10^{-1}, 10^{-2}, 10^{-3}, 10^{-4}$ dyne~cm$^{-2}$
superimposed on color-coded ionization parameter $\log \xi(r,\theta)$
that is normalized to its maximum value. (b) A schematic
description of stratified MHD-driven wind structure manifesting WAs
and UFOs (not to scale) with column distribution (similar to Fig.~8
in K12 and Fig.~5 in T12c). The arrows indicate a mutual interaction between
inner relativistic jets and the MHD winds at their interface. }
\label{fig:fig5}
\end{figure}

%
Besides the above connection between WAs and UFOs we would further
like to point out  that the highly ionized section of the wind
near its axis provides thermodynamic characteristics consistent with
radio jet properties. To this end, we show in
Figure~\ref{fig:fig5}a (in linear space again), the innermost ($r
\lesssim 800 r_o$) structure of model (A): the contours of wind
poloidal velocity $v_p/v_o$ are given by the solid curves (for
$v_p/c=0.2, 0.1, 0.05, 0.03, 0.01$ from innermost to outermost)
while those of the wind pressure by the dashed ones (for $p_{\rm
UFO}=10^{-1}, 10^{-2}, 10^{-3}, 10^{-4}$ dyne~cm$^{-2}$ from bottom to top)
and are superimposed on the color coded ionization structure $\log
\xi(r,\theta)$, normalized to its maximum value.
In this region, the pressure $p(200r_o,40\degr) \sim 0.012$
dyne~cm$^{-2}$ and velocity $v_p/c \gtrsim 0.01$ are consistent with
those obtained for the UFOs of the radio galaxy 3C~111 $p \gtrsim
0.001-0.01$ dyne~cm$^{-2}$ (T12b), which were also shown to be
compatible with the pressure of radio jets in the same region.
%
%
Because of their tighter collimation, we would therefore associate
models (A) and (D) with the structure of radio loud Seyferts.
Finally, a choice between these two models will rely on the
dependence of the value of $N_H$ on the ionization parameter $\xi$,
assuming that the entire X-ray spectrum is produced by a point-like
source, i.e. a source whose extent does not depend on the X-ray
energy.

\section{Summary \& Discussion}



We have presented above the 2D ionization structure of the MHD
winds which our previous works we have associated with the WAs of
AGNs. We have argued further that the same winds are also compatible
with the more recently discovered UFOs and that in fact these
features are different facets of the same underlying phenomenon, which
apparently also occurs in radio galaxies, like 3C~111. We have
also argued that the radio/X-ray observations of this later object
imply pressure equilibrium between its radio jet emitting plasma and
the plasma associated with its UFOs, suggesting that the former
occupies, in pressure equilibrium, the near-axis region of the
self-similar UFO-WA flows of our models. 

The main results of the present
work are encapsulated in Figures 1 and 4: they present the extreme cases of wind collimation geometry
compatible with our assumptions for two different values of the
wind radial density dependence. We found that, while the density
profile does affect the wind collimation, the latter is determined
mainly by the value of their specific angular momentum $H_o$, with
higher values producing significantly less collimation than lower
ones.
The issue of jet/MHD wind collimation is of interest in studies of
AGN statistics, considering that the outer regions of these winds could 
play the role of the AGN unification ``torus" \citep{KK94}. 
Thus, higher collimation implies larger fraction of sources with
X-ray absorption features. A more collimated flow, because of
less ambient matter entrainment, is more likely to produce a well
defined, large scale narrow jet  and therefore of higher velocity.
Similarly, less collimated flows are likely to entrain more matter
leading to lower velocity jets.
%
The point of the present work is to note that, while it is difficult
to image the regions where these flows originate and collimate, one
can obtain their physical conditions (column, velocity) from
spectroscopic X-ray observations and also its collimation properties
from the statistics of AGNs with a given amount of obscuration
\citep{Malizia09, Terashima10} and their relation to the direction
of a larger scale radio jet. A resolution of these issues, as well
as those of systematics such as the obscuration dependence on
luminosity \citep{Tueller08, Burlon11}, require further observations
and modeling \citep{Proga03,Everett05,K12}. However, we believe that
large scale MHD winds will form the basis for an account of such
global properties.  Figure~\ref{fig:fig5}b illustrates a proposed manifestation of  apparently distinct absorbers in the MHD wind discussed in this study. While it is beyond the scope of this work, we also note in a global viewpoint that the radial extent of magnetized winds is coupled to the disk magnetization \citep[][]{Murphy10}.

In addition, there is an issue of the radial density dependence of
these winds. As we discussed in the introduction, the main
motivating factor of our previous work, have been the observations
of \citet{B09} that indicate $N_H \propto \xi^s$ where $s \simeq 0$ and
imply a LoS density profile $n(r) \propto r^{-(2s+1)/(s+1)}, ~ 0
<s<1 $ with most values closer to $s \simeq 0$ [the parameter $s$ is
related to that $q$ of Eq. (\ref{eq:B}) by $2q = (s+2)/(s+1)$]. As
we noted this scaling suggests correlations between the column
$N_H$, the velocity $v$ and the ionization parameter $\xi$ of
specific transitions at a given object (for instance $N_H \propto
v^{2s/(s+1)}, ~v \propto \xi^{(s+1)/2})$. More recently,
observations, mainly of UFOs but also certain WA transitions, in an
ensemble of (rather than a single) AGNs
(\citealt{Tombesi10b,Tombesi12a},~T12b), have shown on the
aggregate, a preference for the value $s \simeq 1$ implying  a LoS
density profile closer to  $n(r) \propto r^{-3/2}$. At this point it
would be of interest to note that models that include a combined
accretion -- outflow approach to the problem \citep{Ferreira97,FC04}
are more restricted in their choice of parameters than those
presented above and do provide additional restrictions to the values
of $s$. So, ``cold" flows \citep{Ferreira97} are restricted to
density profiles very close to $n(r) \propto r^{-3/2}$, while models
with ``entropy injection" near the \Alfven\ surface \citep{FC04} can
yield density profiles close to $n(r) \propto r^{-1}$. At
this point we will reiterate that X-ray spectroscopy observations
could provide the distinction between these extreme cases thus
establishing a probe of the accretion -- outflow  physics of these
systems.

Another issue with the solutions we present is their reliance
on self-similarity. The utility of this technique is not only
mathematical, i.e. in allowing one to obtain solutions of the MHD
equations; it is of importance because it provides solutions that
span a multiple decades in radius. Solutions of this type are
demanded by the data, i.e. the large range in ionization parameter
spanned by the observed X-ray transitions, as discussed in \S 1. Whether the simple functional form of the model 
parameters given above extends over the entire domain of validity of
these solutions is an issue that should be answered observationally.
Deviations from a single power law would indicate the presence of
additional physics that will have to be included in such
calculations.

By their nature, self-similar solutions cannot cover the region near
the flow symmetry axis and they must also terminate at some distance
from the origin. As we have noted above, the near-axis high latitude
region is much less dense and much more highly ionized than the
lower latitude ones, so from the interpreting observed point of view of
absorption features which is the main thrust of the present paper, it is
not of significance. It is certainly of significance mathematically and for impacting  the physics of jets, as the magnetic
fields of this  region are likely to thread the BH event horizon. Pressure
balance requires that the pressure there match that at larger radii
where  the self-similar solutions are valid. Possibly this is the
pressure of relativistic particles that give rise to the radio jets
observed. Such models that combine a self-similar disk-wind solution
with an MHD flow of limited lateral extent (i.e. one in which the
magnetic field threads a star or a BH), have recently
appeared in the literature \citep{Mats08,Mats09,PorthFendt10}. 
Again, it is very conceivable that the winds may consist of {\it two} components; an axial collimated {\it jet} of low matter density probably originating from the BH itself surrounded by an extended {\it disk-wind} \citep[e.g.][]{Ferreira10} also similar to our view depicted in Figure~\ref{fig:fig5}b.
It is
of interest to note that the last of the above references finds
flows with opening angles between 4\degr and 6\degr, not very
different of those of models (A) and (D).

At large radii, these solutions cannot be trusted beyond an
equatorial distance larger than that of the underlying disk. This
distance is not known. However, it cannot be larger than the black
hole influence radius, $R_B \sim M_{\rm BH}/\sigma^2$, i.e. the distance
at which the disk Keplerian velocity matches the velocity dispersion
$\sigma = \langle v^2 \rangle^{1/2}$ of the AGN surrounding
spheroid. However, given the relation between $M_{\rm BH}$ and $\sigma$,
namely ($M_{\rm BH}/10^8 M_{\odot}) \sim (\sigma/ 200 \,{\rm km \,
s}^{-1})^4$ \citep[][]{Ferrarese00}, this distance is of order $R_B
\sim 10^{19} (M_{\rm BH}/10^8 M_{\odot})^{-1/2} \sim 10^{19} M_8^{-1/2}$
cm. It is of interest to note that absorption feature observations,
in conjunction with our models, could provide an independent
estimate of this distance and therefore an independent measure of
the $M - \sigma$ relation.

Finally, one should bear in mind that density profiles
$n(r) \propto r^{-\alpha}$, produce winds whose mass flux increases
with distance for $\alpha \le 3/2$. Applying the profile with
$\alpha =1$ and $\dot m =0.1$ used in our models to an accreting
BH of $M = 10^8 \, M_{\odot}$ we obtain a total mass flux of
%
%
\begin{eqnarray}
\dot{M}_{\rm out}^{\rm (global)}  &\equiv& m_p \int_{z=z_c} d^2 x ~ n(r,\theta) v_z
\sim \frac{4 \sqrt{2} G^2}{c^5} F B_o^2 M^2 r_{\rm out}^{1/2} \nonumber \\
&\sim& 6.5 \Msun \left(\frac{F}{0.1}\right) \left(\frac{B_o}{10^4\textrm{G}}\right)^2
\left(\frac{M}{10^8\Msun}\right)^2 \left(\frac{r_{\rm out}}{10^6 r_o}\right)^{1/2}
~ \textrm{year}^{-1} \ , \label{eq:Mdot}
\end{eqnarray}
a value much higher than that needed to power its bolometric
luminosity of $\sim 10^{45}$ erg/s. 
The corresponding (integrated) kinetic
power\footnote[3]{Also see K12 for the discussion of
radial-dependence of the wind quantities.} is given by
$\dot{E}_{\rm out}^{\rm (local)}  \equiv \dot{M}^{\rm (local)} v_{\rm out}^2 \propto r^{-1/2}$.
Assuming $r_{\rm in}/r_o \sim 100$ this yields
$\dot{E}_{\rm out}^{\rm (global)} \sim 10^{44}$
erg~s$^{-1}$ that is indeed consistent
with the estimated UFO power in 3C~111
(\citealt{Tombesi10b,Tombesi12a},~T12b) providing a large impact
on AGN feedback process at large scales.
%
%
On the other hand, the cumulative power of UV/soft X-ray WAs may be able to reach this level as well \citep[e.g.][]{CK12}.
It is interesting to note that
outflow rates, estimated by X-ray spectroscopy in BH binaries 
\citep{Lee09,Neilsen12}, were found to be much larger than the
accretion rates needed to power their X-ray luminosity. 

On a closing note, although in this work we primarily focused on physical conditions
and a {\it global} geometrical structure of MHD winds, we are currently simulating
 the corresponding spectral
signatures of the absorption lines within this model (Fukumura et
al.~2013, in prep) .  To constrain individual wind parameters
(such as $F_o, H_o$ and their dependency on $\dot m$ and $\theta$),
however, one may require numerical calculations of X-ray absorption feature
with sets of MHD parameters and perform a $\chi^2$-test in comparison with data; i.e. observationally constrain the
wind parameters that would otherwise be  inaccessible. While it is still
challenging to constrain all the wind variables, one can still hope
to restrict the range of wind parameters allowed by spectrum
analysis.
The capabilities of the upcoming mission {\it Astro-H} should be instrumental in performing these insightful observations.

\acknowledgments

KF thanks an anonymous referee for improving  the manuscript and M.
Nakamura for his inspiring comments on the draft manuscript.  EB is
supported by grants from the Israel Science Foundation and from
Israel's Ministry of Science and Technology.

\section*{Appendix: A Brief Comparison of Our Wind Parameters with BP82 Winds}

We have computed and examined four magnetized wind solutions to
steady-state, axisymmetric ideal MHD equations under Newtonian
gravity (as in CL94, FKCB10a, FKCB10b and K12) in the context of its potential
role of outflow physics. Generic properties of MHD outflows
considered in this study include mass-loading and angular
momentum as well as the global density profile. We suggest that smaller angular
momentum tends to help collimate the outflow poloidal structure
while a lower plasma-to-magnetic flux ratio generally leads to a more
efficient wind acceleration by the action of a large-scale magnetic
fields (at least predominantly relative to other mechanisms such as
radiation pressure).

Here, we would like to show a viability of the sets of our wind parameters by recalling some of the fundamental definitions of the conserve quantities. As discussed in CL94, one can express $F_o$ in terms of some ``known" quantities from equation~(\ref{eq:F}) combined with
%
\begin{eqnarray}
F(\Psi) = r_\Psi^{q-3/2} F_o \frac{B_o}{v_o} \ ,
\end{eqnarray}
to obtain
\begin{eqnarray}
F_o = 4 \pi \rho \frac{v_p}{B_p} \frac{v_o}{B_o} \sim 10^{11} (0.01 \times (0.7c)^2)/(10^4)^2 \sim \mathcal{O}(10^{-2})  \ .
\end{eqnarray}
assuming a typical set of values (for AGNs) of $n_o \sim 10^{10}$ cm$^{-3}$ (Crenshaw et al.~2003 and references therein), $B_p \sim B_o \sim 10^4$ G, and $v_p \sim 0.01 v_o$ at the innermost launching radius. This value is in consistence with our chosen values. For the rest of the parameters we (this paper and CL94) are also convinced of consistency with our parameter sets in comparison with BP82 winds as also discussed in CL94; i.e. one can make a one-to-one mapping between BP82 wind parameters and ours by noting that
\begin{eqnarray}
\lambda \leftrightarrow -H_o/F_o  \ ,
\end{eqnarray}
where BP82's characteristic value of $\lambda=30$ leads to $-H_o \sim \mathcal{O}(1)$. From our boundary condition at the disk surface expressed in equation~(\ref{eq:Omega}) (see also CL94) one  finds
\begin{eqnarray}
\Omega_o = 1-(F_o + H_o) V_{z,0} \sim \mathcal{O}(1)  \ ,
\end{eqnarray}
where we have used $F_o$ and $H_o$ above together with $V_{z,0} \sim 0.01$. Therefore, we are confident in our parameter choice.

\end{document}